\documentclass[12pt]{iopart}
\usepackage{harvard}
\usepackage{bibtexlogo}
\usepackage{graphicx}
\usepackage{iopams}
\usepackage{subfigure,verbatim}

\newcommand{\unit}[1]{\nobreak{\mathrm{\;#1}}} 

\newcommand{\bmath}[1]{\mbox{\boldmath{$#1$}}}

\def\comp{\,c/\omega_{\rm p}}

\def\ompt{\omega_{\rm p}t}

\newcommand{\gb}[1]{\gamma\beta_{\rm {#1}}}

\newcommand{\eq}[1]{eq.~(\ref{eq:#1})}
\newcommand{\fig}[1]{Fig.~\ref{fig:#1}}
\newcommand{\fign}[1]{\ref{fig:#1}}
\newcommand{\tit}[1]{\textit{#1}}
\newcommand{\be}{\begin{eqnarray}}
\newcommand{\ee}{\end{eqnarray}}
\newcommand{\citet}[1]{\citeasnoun{#1}}
\newcommand{\citep}[1]{\cite{#1}}

\begin{document}

\title[Shock-driven reconnection in relativistic striped winds]{Particle-in-cell simulations of shock-driven reconnection in relativistic striped winds}

\author{Lorenzo Sironi$^1$ and Anatoly Spitkovsky$^2$}

\address{$^1$ Institute for Theory and Computation,
Harvard-Smithsonian Center for Astrophysics,
60 Garden St.,
Cambridge, MA 02138, USA\\
$^2$ Department of Astrophysical Sciences, Princeton University, 4 Ivy Lane,\\
Princeton, NJ 08544-1001,
USA\\}
\eads{\mailto{lsironi@cfa.harvard.edu}, \mailto{anatoly@astro.princeton.edu}}
\begin{abstract}
By means of two- and three-dimensional particle-in-cell simulations, we investigate the process of driven magnetic reconnection at the termination shock of relativistic striped flows. In pulsar winds and in magnetar-powered relativistic jets, the flow consists of  stripes of alternating magnetic field polarity, separated by current sheets of hot plasma. At the wind termination shock, the flow compresses and the alternating fields annihilate by driven magnetic reconnection. Irrespective of the stripe wavelength $\lambda$ or the wind magnetization $\sigma$ (in the regime $\sigma\gg1$ of magnetically-dominated flows), shock-driven reconnection transfers all the magnetic energy of alternating fields to the particles, whose average Lorentz factor increases by a factor of $\sigma$ with respect to the pre-shock value. In the limit $\lambda/(r_L\sigma)\gg1$, where $r_L$ is the relativistic Larmor radius in the wind, the post-shock particle spectrum approaches a flat power-law tail with slope around $-1.5$, populated by particles accelerated by the reconnection electric field. The presence of a current-aligned ``guide'' magnetic field suppresses the acceleration of particles only when the guide field is stronger than the alternating component. Our findings place important constraints on the models of non-thermal radiation from Pulsar Wind Nebulae and relativistic jets.
\end{abstract}

\vspace{-0.35in}
\submitto{Computational Science \& Discovery -- Special issue on selected research from the $22^{nd}$ International Conference on Numerical Simulation of Plasmas -- ICNSP 2011}

\section{Introduction}\label{sec:intro4}
Pulsar Wind Nebulae (PWNe) are bubbles of synchrotron-emitting plasma powered by the relativistic wind  of rotation-powered pulsars. Their broadband spectrum is produced by a nonthermal population of electron-positron pairs (hereafter, simply ``electrons''), presumably accelerated at the wind termination shock, where the momentum flux of the relativistic pulsar wind is balanced by the confining pressure of the nebula. The observed radio spectrum \citeaffixed{bietenholz_97}{$F_{\nu}\propto\nu^{-0.25}$ for the Crab Nebula;}  implies a distribution of emitting electrons of the form $dN/dE\propto E^{-p}$, with a  flat slope $p\sim1.5$. To explain the radio through optical emission of the Crab Nebula, the power law of shock-accelerated electrons should span at least three decades in particle energy \cite{lyubarsky_03}. Flat electron spectra with slopes $p<2$ below GeV energies are also required to model the radio emission of hotspots in radio galaxies \citep{stawarz_07} and X-ray observations of luminous blazar sources \citep{sikora_09}. 

A flat power-law spectrum with index $1<p<2$ is not  expected from the standard theory of Fermi acceleration in relativistic shocks, which normally yields slopes $p>2$ \cite{achterberg_01,keshet_waxman_05,sironi_spitkovsky_09,sironi_spitkovsky_11a}. An acceleration mechanism capable of generating flat spectra in PWNe was discussed by \citet{lyubarsky_03} and studied by \citet{sironi_spitkovsky_11b}, under the assumption that the flow upstream of the termination shock consists of alternating stripes of opposite magnetic polarity, separated by current sheets of hot plasma (from now on, a ``striped wind''). For obliquely-rotating pulsars, this is the configuration expected around the equatorial plane of the wind, where the sign of the toroidal field alternates with the pulsar period. A similar geometry is invoked in the proto-magnetar model of gamma-ray bursts, if the striped structure of the equatorial wind is preserved when the flow gets redirected along the polar jet \citep{metzger_11}. When passing through a shock (e.g., the termination shock of pulsar winds), the flow compresses and the alternating fields annihilate by shock-driven reconnection, which may produce the flat electron spectrum required by the observations.  

The physics of magnetic reconnection can be captured self-consistently only by means of multi-dimensional particle-in-cell (PIC) simulations. Fully-kinetic PIC simulations provide a powerful tool to explore the microphysics of collisionless plasmas, since they can capture from first principles the fundamental interplay between charged particles and electromagnetic fields. In the context of relativistic magnetic reconnection in pair plasmas, most studies have explored the process of undriven reconnection \citep{zenitani_01,jaroschek_04}, where field annihilation is initiated by a transient seed perturbation to an otherwise stable current sheet. As discussed above, this is not the setup expected at the  termination shock of relativistic striped winds. There, it is the shock-compression of the flow that steadily drives regions of opposite magnetic field polarity toward each other, causing reconnection. Recent experimental and numerical studies of non-relativistic plasmas have shown that driven reconnection is much faster than the undriven process \citep{fox_11}.

In this work, we explore via multi-dimensional fully-kinetic PIC simulations the physics of driven reconnection at the termination shock of a striped relativistic electron-positron wind. We find that the alternating fields are completely dissipated upon compression by the shock, and their energy is transferred to the particles, regardless of the properties of the flow.  Broad particle spectra with  slopes $1<p<2$ are a common by-product of shock-driven reconnection, but the extent of the power-law tail depends on the wind magnetization and the stripe wavelength.

\section{Simulation Setup}\label{sec:setup}
We use the 3D electromagnetic PIC code TRISTAN-MP \citep{buneman_93,spitkovsky_05} to study the termination shock of a relativistic striped wind. PIC codes can model astrophysical plasmas in the most fundamental way, as a collection of charged macro-particles that are moved by integrating the Lorentz force. The currents deposited by the macro-particles on the computational grid are then used to advance the electromagnetic fields via Maxwell's equations. The loop is closed self-consistently by extrapolating the fields to the macro-particle locations, where the Lorentz force is computed. 

\begin{figure}[tbp]
\begin{center}
\includegraphics[width=0.6\textwidth]{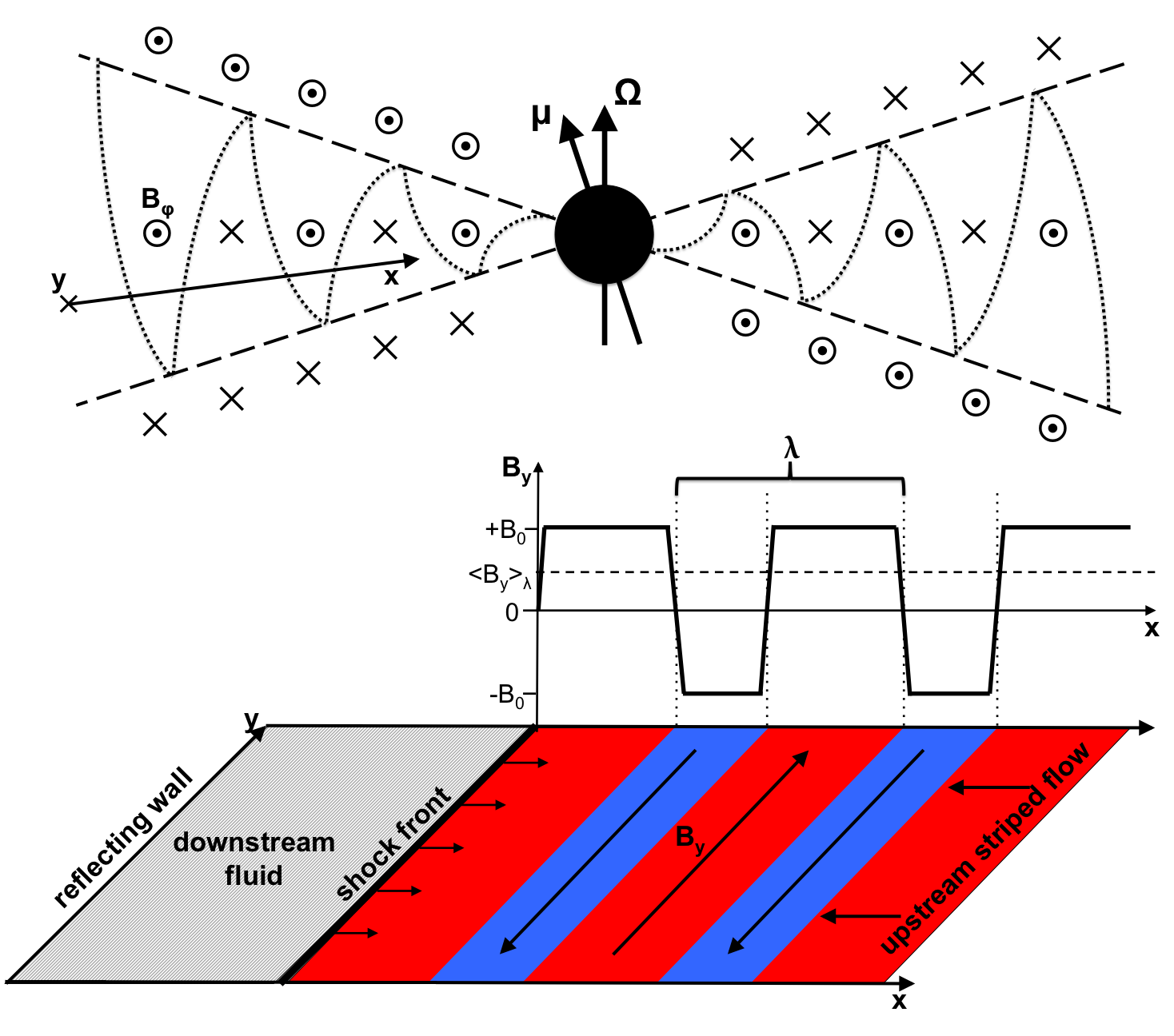}
\caption{Upper panel: poloidal structure of the striped pulsar wind, according to the solution by \citet{bogovalov_99}. The arrows denote the pulsar rotational axis (along $\bmath{\Omega}$, vertical) and magnetic axis (along $\bmath{\mu}$, inclined). Within the equatorial wedge bounded by the dashed lines, the wind consists of toroidal stripes of alternating polarity, separated by current sheets (dotted lines).  Lower panel: simulation geometry. For 2D runs, the simulation domain is in the $xy$ plane, oriented as shown in the upper panel. The incoming flow propagates along $-\bmath{\hat{x}}$, and the shock moves away from the reflecting wall (located at $x=0$) toward $+\bmath{\hat{x}}$. The magnetic field lies in the simulation plane along the $y$ direction, and its polarity alternates with wavelength $\lambda$. A net stripe-averaged field $\langle B_y\rangle_\lambda>0$ is set up by choosing red stripes (where $B_y=+B_0$) wider than the blue stripes  (where $B_y=-B_0$). For the pulsar wind sketched above, $\langle B_y\rangle_\lambda>0$ is realized below the equatorial plane.}
\label{fig:simplane}
\end{center}
\end{figure}

We mainly utilize 2D computational domains in the $xy$ plane, with periodic boundary conditions in the $y$ direction. In \S\ref{app:3d} we compare 2D and 3D simulations and show that 2D runs can capture most of the relevant physics. For both 2D and 3D domains, all three components of particle velocities and electromagnetic fields are tracked. The shock is set up by reflecting a magnetized electron-positron  flow from a conducting wall located at $x = 0$ (\fig{simplane}). The interaction between the incoming beam (that propagates along $-\bmath{\hat{x}}$) and the reflected beam triggers the formation of a shock, which moves away from the wall along $+\bmath{\hat{x}}$ (\fig{simplane}). The simulation is performed in the ``wall'' frame, where the post-shock plasma is at rest. The incoming stream is injected with bulk Lorentz factor $\gamma_0=15$, but we have explored a wide range of $\gamma_0$, finding the same results, modulo an overall shift in the energy scale \citep{sironi_spitkovsky_11b}.

The upstream flow carries a strong magnetic field of intensity $B_0$, oriented along the $y$ direction. The choice to initialize the field in the simulation plane (i.e., oriented along $y$, as opposed to $z$) is motivated by the agreement between our 2D simulations with in-plane fields and 3D experiments, as we show in \S\ref{app:3d}. The field alternates with wavelength $\lambda$. We vary $\lambda$ from $\lambda=20\comp$ up to $\lambda=1280\comp$, where $c/\omega_{\rm p}\equiv \sqrt{\gamma_0 m c^2/ 4\pi e^2 n_{c0}}$ is the relativistic plasma skin depth in the wind. Here, $n_{c0}$ is the density of particles  in the wind. The thickness of each current sheet is a few plasma skin depths (wide enough so that the process of undriven reconnection is suppressed in our simulations), but our results are not sensitive to this parameter, provided that the contribution of current sheets to the particle and energy flux in the wind is negligible (see Appendix A). We parameterize the field strength $B_0$ via the so-called magnetization parameter $\sigma\equiv B_{0}^2/4 \pi \gamma_0 m n_{c0} c^2$. We vary $\sigma$ from 10 up to 100. Although the magnetic field strength in the  wind is always $B_0$, the wavelength-averaged field $\langle B_y\rangle_\lambda$ can vary from zero up to $B_0$, depending on the relative widths of the regions of  positive and negative field (lower panel in \fig{simplane}). In  pulsar winds, one expects $\langle B_y\rangle_\lambda=0$ only in the equatorial plane (where the stripes are symmetric), whereas $|\langle B_y\rangle_\lambda|/B_0\rightarrow1$ at high latitudes (upper panel in \fig{simplane}). We choose $\alpha\equiv2\langle B_y\rangle_\lambda/(B_0+|\langle B_y\rangle_\lambda|)=0.1$ (corresponding to $\langle B_y\rangle_\lambda/B_0\simeq0.05$), to represent the behavior of the shock in the vicinity of the equatorial plane. \citet{sironi_spitkovsky_11b} have explored the full range $0\leq\alpha\leq1$. We remark that a net field $\langle B_y\rangle_\lambda\neq0$ does not play the role of a guide field, which in our geometry would correspond to a uniform component of magnetic field in the $z$ direction (aligned with the particle velocity in the current sheets). The guide field is expected to be negligible in pulsar winds, but it may be dominant in gamma-ray burst and blazar jets. We explore the effects of a guide field on the physics of shock-driven reconnection in \S\ref{sec:guide}.

We choose the relativistic electron (or positron) skin depth in the cold wind $c/\omega_{\rm p}$ such that the smallest scale in the system, which for $\sigma>1$ is the relativistic Larmor radius $r_{L}=(c/\omega_{\rm p})/\sqrt{\sigma}$, is resolved with at least a few computational cells.\footnote{The Debye length of the cold upstream wind is also resolved with a few computational cells, to avoid the finite grid instability and artificial heating.} We usually choose $r_L=3$ cells, but we have verified that our results do not change when doubling the resolution. Our computational domain is typically $\sim400\comp$ wide (along the $y$ direction). As we show in Appendix B, a large box is essential for the consistency of our findings.  Each computational cell is initialized with two electrons and two positrons, but we have performed limited experiments with a larger number of particles per cell (up to 32 per species), obtaining essentially the same results.

\section{Shock Structure}\label{sec:shock}
In this section, we investigate the steady-state structure of a relativistic shock propagating in a striped electron-positron wind. We adopt $\gamma_0=15$, $\sigma=10$, and $\langle B_y\rangle_\lambda/B_0\simeq0.05$ (corresponding to $\alpha=0.1$) as our fiducial values for the upstream bulk Lorentz factor, magnetization, and strength of the stripe-averaged field. To clarify the physics of shock-driven reconnection, we employ a relatively large value for the stripe wavelength ($\lambda=320\comp$ or $\lambda=640\comp$), and we neglect the guide field. In \S\ref{sec:cond} we discuss the dependence of our results on the stripe wavelength and the wind magnetization, and in \S\ref{sec:guide} we comment on the effects of a nonzero guide field. 

\begin{figure}[tbp]
\begin{center}
\includegraphics[width=0.65\textwidth]{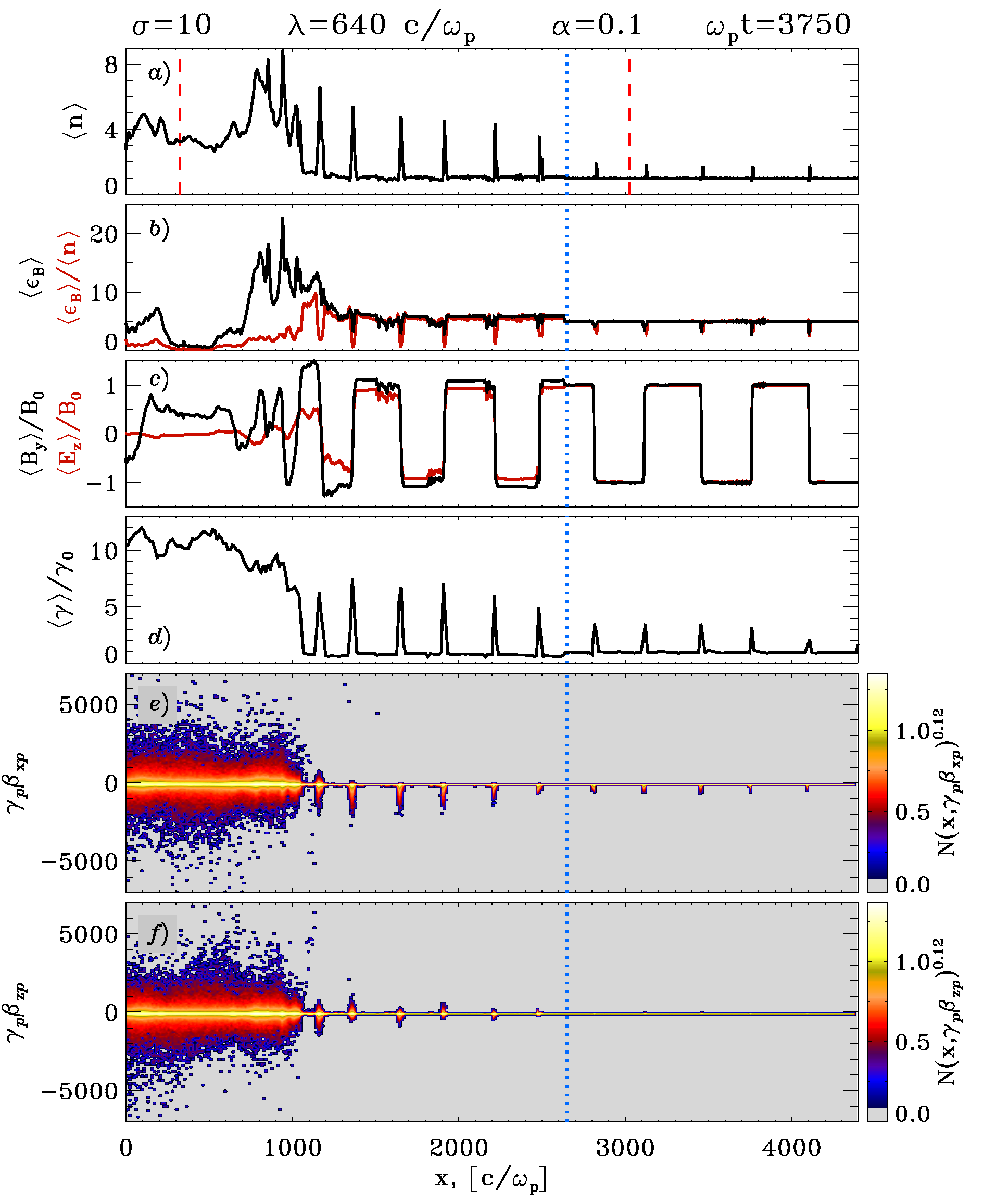}
\caption{Internal structure of the flow at $\ompt=3750$, for $\lambda=640\comp$, $\sigma=10$, and $\alpha=0.1$. See the supplementary material for an associate movie (\textsf{1d\_shock\_structure.mov}) following the complete evolution of the shock. At $\ompt=3750$, the hydrodynamic shock is located at $x\simeq1000\comp$, and the MHD shock at $x\simeq2600\comp$, as indicated by the vertical dotted blue line. As a function of the longitudinal coordinate $x$, we plot: (a) $y$-averaged particle number density $n$, in units of the upstream value; (b) $y$-averaged magnetic energy fraction $\epsilon_B\equiv B^2/8 \pi \gamma_0 m n_{c0} c^2$ (black line), and $y$-averaged magnetic energy per particle $\epsilon_B/n$ (red line); (c) $y$-averaged magnetic field $B_y$ (black line) and electric field $E_z$ (red line), normalized to the upstream magnetic field $B_{0}$; (d) mean kinetic energy per particle, in units of the bulk energy at injection; (e) (f) $x-\gb{x}$ and $x-\gb{z}$ positron phase space.}
\label{fig:fluidtot}
\end{center}
\end{figure}

\begin{figure*}[tbp]
\begin{center}
\includegraphics[width=0.85\textwidth]{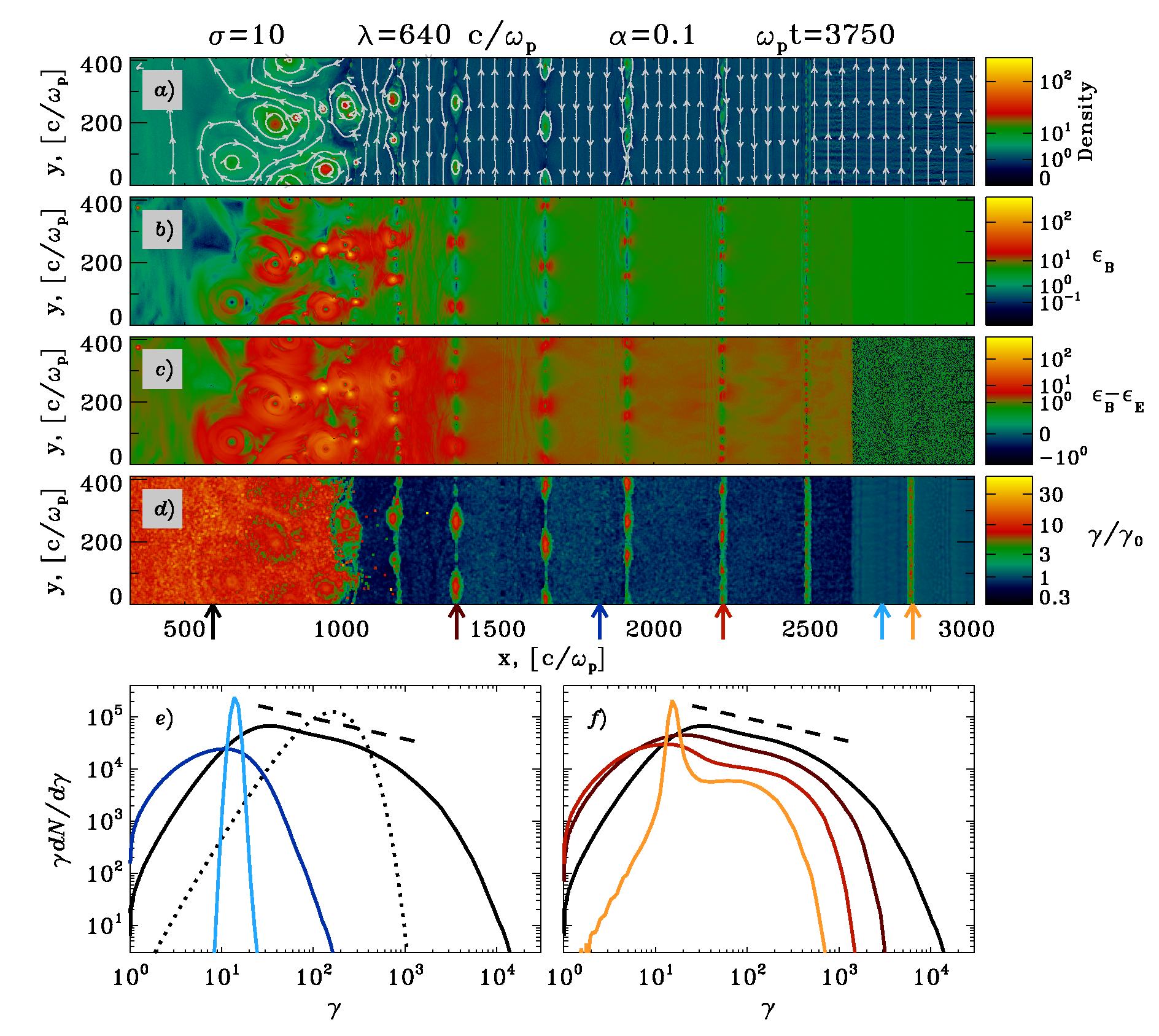}
\caption{Internal structure of the flow at $\ompt=3750$, for $\lambda=640\comp$, $\sigma=10$, and $\alpha=0.1$, zooming in on a region around the shock, as delimited by the vertical dashed red lines in \fig{fluidtot}(a). See the supplementary material for an associate movie (\textsf{2d\_shock\_structure.mov})  following the complete evolution of the shock. At $\ompt=3750$, the hydrodynamic shock is located at $x\simeq1000\comp$, and the fast MHD shock at $x\simeq2600\comp$.  We show the 2D plots of: (a) particle number density, with contours showing the magnetic field lines; (b) magnetic energy fraction $\epsilon_B$; (c) $\epsilon_B-\epsilon_E$, where $\epsilon_E\equiv E^2/8 \pi \gamma_0 m n_{c0} c^2$; (d) mean kinetic energy per particle. Panels (e) and (f) show the particle energy spectra, respectively outside (e) or inside (f) of current sheets. The color of each spectrum matches the color of the corresponding arrow at the bottom of panel (d), showing where the spectrum is computed. The dotted line in panel (e) is a Maxwellian distribution with the same average energy as the downstream particles; the dashed lines in panels (e) and (f) indicate a power law with slope $p=1.4$.}
\label{fig:fluidsh}
\end{center}
\end{figure*}

The longitudinal profile of the shock transition region, along the direction $x$ of shock propagation, is presented in \fig{fluidtot}, whereas \fig{fluidsh} shows the flow properties in the $xy$ plane of the simulation, zooming in on a region around the shock (as delimited by the vertical dashed red lines in \fig{fluidtot}(a)). The steady-state structure of the flow shows the presence of two shocks. The main shock (which we shall call the ``hydrodynamic shock'' from now on, for reasons that will become clear below) corresponds to the jump in density occurring at $x\simeq1000\comp$ in \fig{fluidtot}(a). At $x\simeq2600\comp$, well ahead of the hydrodynamic shock, the incoming flow crosses a fast magnetohydrodynamic (MHD) shock, whose location is highlighted in \fig{fluidtot} by a vertical dotted blue line. 

At the fast shock, the wind is decelerated and compressed. 
The compression induced by the passage of the fast shock through a given current sheet triggers the onset of magnetic reconnection. Small-scale islands develop inside the current sheet as a result of the tearing-mode instability. The growth and evolution of reconnection islands downstream from the fast shock ($x\lesssim2600\comp$) is clearly shown in the 2D plots of density and magnetic energy of \fig{fluidsh}(a) and (b), respectively. A magnetic X-point exists in between each pair of neighboring islands, where field lines of opposite polarity break and reconnect, and the magnetic energy is transferred to the particles. Particles flow into the X-points and are accelerated by the reconnection electric field (see the green and blue regions in the 2D plot of $\epsilon_B-\epsilon_E$ of \fig{fluidsh}(c)), while being advected by the reconnected magnetic field into the closest island. Magnetic islands act as reservoirs of particles (\fig{fluidsh}(a)) and particle energy (the high-energy regions of \fig{fluidsh}(d), colored in red, match very well the overdense islands of \fig{fluidsh}(a)). As the flow recedes from the fast shock, the small-scale islands created inside each current sheet grow and coalesce to form bigger islands. This temporal evolution corresponds, in the snapshot of \fig{fluidsh}, to a pattern of bigger (and so, fewer) islands for current sheets that are farther behind the fast shock. Reconnection proceeds with distance downstream from the fast shock, up to the point where reconnection islands fill the entire region between neighboring current sheets. Now, the striped structure of the flow is erased, and a hydrodynamic shock forms (located at $x\simeq1000\comp$ in Figs.~\fign{fluidtot} and \fign{fluidsh}). 

Behind the hydrodynamic shock, the flow comes to rest (in the wall frame of our simulations), and the particle distribution is isotropic in three dimensions (compare longitudinal and transverse phase spaces in \fig{fluidtot}(e) and (f), respectively). The post-shock number density approaches $n_{\rm d}\simeq4\,n_{\rm u}$ (\fig{fluidtot}(a), where $n_{\rm u}$ is the density ahead of the fast shock), and the shock velocity is $\beta_{\rm sh}\simeq1/3$, in agreement with the jump conditions of a relativistic \tit{unmagnetized} 3D plasma. In fact, most of the energy per particle downstream from the hydrodynamic shock is in kinetic form (as opposed to the dominant electromagnetic component of the incoming striped flow). This becomes clear when comparing the profile of the mean \tit{kinetic} energy per particle (\fig{fluidtot}(d)) with the 1D plot of the mean \tit{magnetic} energy per particle (red line in \fig{fluidtot}(b)).\footnote[2]{We point out that in the incoming flow electric and magnetic fields equally contribute to the energy balance, whereas the red line in \fig{fluidtot}(b) only includes magnetic fields. However, downstream from the hydrodynamic shock, the flow is at rest (in the simulation frame), and electric fields vanish (see the red line in \fig{fluidtot}(c)).} Behind the hydrodynamic shock, the average kinetic energy per particle reaches $\langle \gamma\rangle/\gamma_0\simeq\sigma+1\simeq11$, as expected in the case of full dissipation of magnetic fields. Correspondingly, the mean magnetic energy per particle drops to zero (red line in \fig{fluidtot}(b)). Behind the shock, the only residual magnetic field comes from  shock-compression of the stripe-averaged field $\langle B_{\rm y}\rangle_\lambda$ (this is best seen in experiments with a smaller stripe wavelength).

Finally, we remark that, as apparent in \fig{fluidsh}, the tearing mode instability, which plays an essential role for the dynamics of the flow, can be captured correctly only with multi-dimensional simulations. Indeed, for the parameters employed in this section, the one-dimensional model by \citet{petri_lyubarsky_07} would predict negligible field dissipation, in sharp contrast with our findings. As we show in Appendix B, this clearly emphasizes the importance of multi-dimensional physics for our understanding of shock-driven magnetic reconnection.


\section{Particle Spectrum and Acceleration}\label{sec:accel}
The particle energy spectrum at different locations through the flow, as marked by arrows at the bottom of \fig{fluidsh}(d), is shown in \fig{fluidsh}(e) and (f). Panel (e) shows the particle distribution in the cold wind, whereas panel (f) focuses on the hot plasma within current sheets. Downstream from the hydrodynamic shock, the striped structure is completely erased, and no difference persists between current sheets and cold wind. Here, the particle spectrum (black line in both \fig{fluidsh}(e) and (f)) is in the form of a flat power-law tail (with spectral index $p\simeq1.4$, dashed line) extending from $\gamma_{\rm min}\simeq30$ to $\gamma_{\rm max}\simeq500$, where it cuts off exponentially. For comparison, a 3D Maxwellian with the same average Lorentz factor $\langle\gamma\rangle=\gamma_0(\sigma+1)\simeq165$ is plotted in \fig{fluidsh}(e) as a dotted line, to show that the actual particle spectrum is much broader than a thermal distribution.  By following the flow from the fast to the hydrodynamic shock, we now clarify how such a particle spectrum is generated.

Ahead of the fast shock, the spectrum at $x\simeq2850\comp$  (yellow line in \fig{fluidsh}(f)) shows, in addition to the cold plasma in the striped wind, the hot particles that are initialized in the current sheet. Behind the fast shock (red line for $x\simeq2200\comp$, brown line for $x\simeq1350\comp$), the particle spectrum consists of two components: a low-energy peak, reminiscent of the particle distribution far from current sheets (compare with the spectrum measured in the wind, dark blue line in \fig{fluidsh}(e)); and a high-energy component, that grows in number and extends to higher and higher energies, as the flow propagates toward the hydrodynamic shock. This component is populated by particles that were initially outside the current sheet, and have entered the sheet in the course of the reconnection process. By the time the flow enters the hydrodynamic shock, the current sheet occupies the entire flow, which explains the smooth transition from the spectrum within current sheets to the distribution downstream of the hydrodynamic shock (yellow through black lines in \fig{fluidsh}(f)).

\begin{figure}[tbp]
\begin{center}
\includegraphics[width=0.7\textwidth]{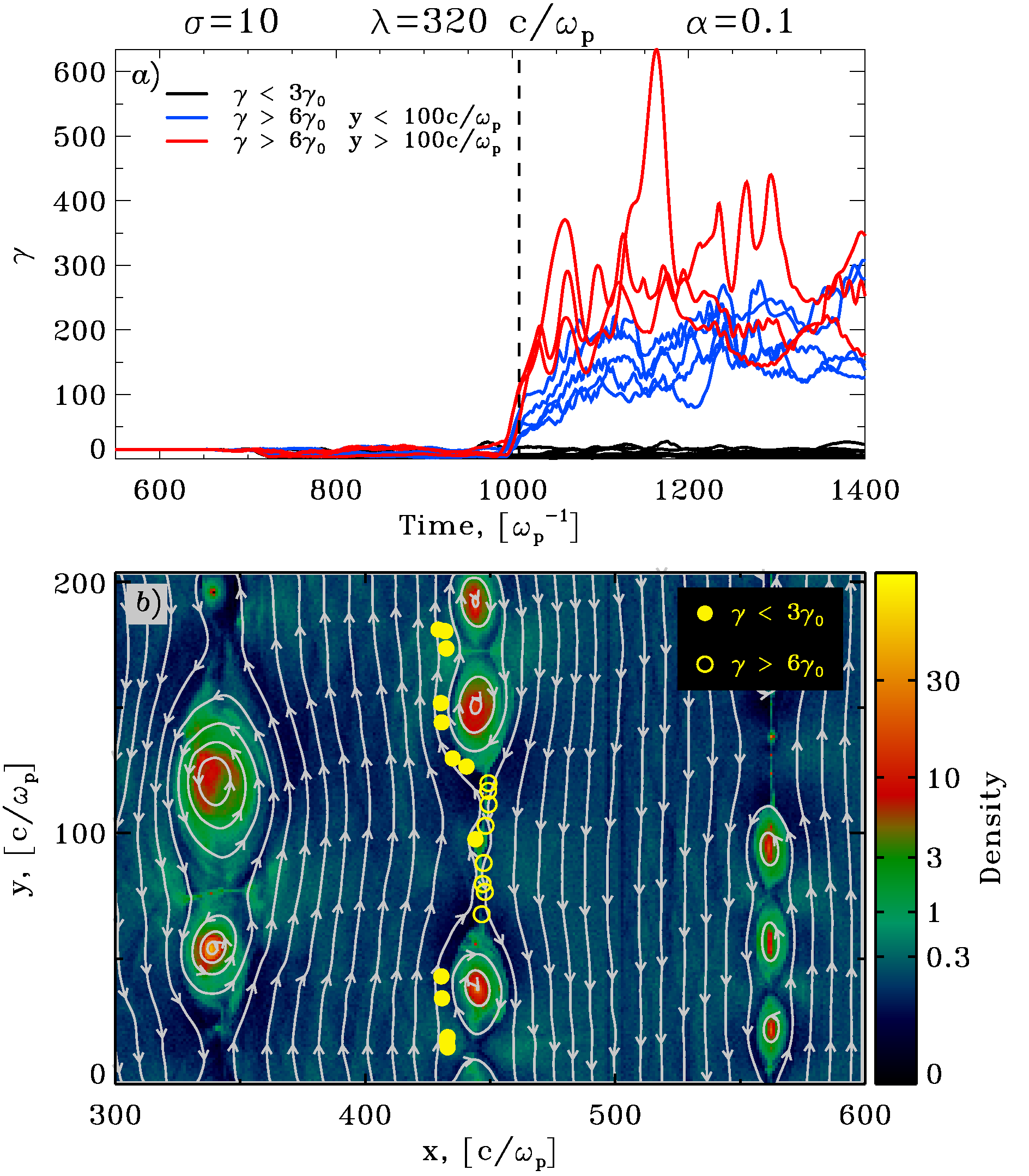}
\caption{Energy evolution of a sample of representative positrons interacting with a current sheet, in a striped flow with $\lambda=320\comp$, $\sigma=10$, and $\alpha=0.1$. See the supplementary material for an associate movie (\textsf{acceleration\_general.mov})  following the complete trajectories of the selected particles. Panel (a): Energy histories of the selected positrons, with lines of different color depending on their final energy (black if $\gamma<3\gamma_0$, blue or red if $\gamma>6\gamma_0$) and their $y$-location at $\ompt=1008$ (blue if $y<100\comp$, red if $y>100\comp$). Panel (b): Magnetic field lines (white contours) superimposed on the 2D plot of particle number density, at time $\ompt=1008$ (marked as a vertical dashed line in panel (a)); depending on the  final particle energy, the locations of the selected positrons are shown as yellow filled (if $\gamma<3\gamma_0$) or open (if $\gamma>6\gamma_0$) circles.}
\label{fig:accel}
\end{center}
\end{figure}

We now investigate in more detail the physics of particle acceleration within current sheets. In \fig{accel}, we follow the trajectories of a representative sample of positrons, extracted from the simulation of a striped wind with $\lambda=320\comp$ (but our conclusions hold for all wavelengths). All the selected positrons are initially in the cold wind (i.e., far from current sheets), at roughly the same $x$-location. They encounter the closest current sheet at $\ompt=1008$, as shown in \fig{accel}(b) together with the corresponding 2D plot of the particle number density. Starting from this time, their histories can diverge significantly, depending on their $y$-location at the moment of interaction with the current sheet. The particles that will end up with relatively low energies ($\gamma<3\gamma_0$, black lines in panel (a)) are found at $\ompt=1008$ exclusively around magnetic islands (their locations are indicated with filled circles in panel (b)). At later times, they will get trapped on the outskirts of the growing islands, without appreciable changes in energy. 

On the contrary, the positrons that will eventually reach high energies ($\gamma>6\gamma_0$, red and blue lines in  \fig{accel}(a)) are concentrated at $\ompt=1008$ in the vicinity of magnetic X-points (the particle locations are shown with open circles in \fig{accel}(b)). Starting from $\ompt=1008$ (vertical dashed line in panel (a)), the energies of these positrons grow simultaneously and explosively (red and blue lines in \fig{accel}(a)), as they are accelerated along $z$ by the reconnection electric field parallel to the X-line \citep{zenitani_01,jaroschek_04,lyubarsky_liverts_08}. Particles pre-accelerated at X-points continue to gain energy by compression \citep{lyubarsky_liverts_08,giannios_10}, as they are advected along $y$ from the X-point into the closest island. No major systematic changes of energy are observed after the particles get trapped within magnetic islands. Different X-points may provide different energy gains, as shown by the red (respectively, blue) curves in panel (a), which refer to high-energy positrons that get accelerated at the X-point above (respectively, below) the small magnetic island in the middle of the current sheet. In summary, the energy gain of a given particle is determined on the one hand by the strength of the reconnection electric field at the X-point, and on the other hand by the time available for acceleration, as the particle flows from the X-point into the closest island. As the flow propagates downstream from the fast shock, the typical distance between an X-point and the closest island increases, allowing particles to be accelerated to higher energies. This explains why the upper cutoff in the particle spectra of \fig{fluidsh}(f) shifts to higher energies for current sheets farther behind the fast shock.

The validity of our conclusions may be questioned if the structure of X-points in a realistic 3D scenario is much different from the 2D picture analyzed here. In particular, in 3D we expect the current sheet to be folded, by effect of the so-called drift-kink instability \citep{zenitani_05,zenitani_07}. The characteristic wavelength of the kink mode in the $xz$ plane will introduce a different length scale, that could potentially compete with the distance between X-points and islands (in the $xy$ plane) in constraining the maximum energy of accelerated particles.  In \S\ref{app:3d} we address this important point, and show that the latter constraint is more restrictive than the former, at least for the parameters explored in this work. It follows that our 2D simulations with in-plane fields can capture very well the main 3D properties of the shock, and in particular the physics of particle acceleration.

\begin{figure}[tbp]
\begin{center}
\includegraphics[width=\textwidth]{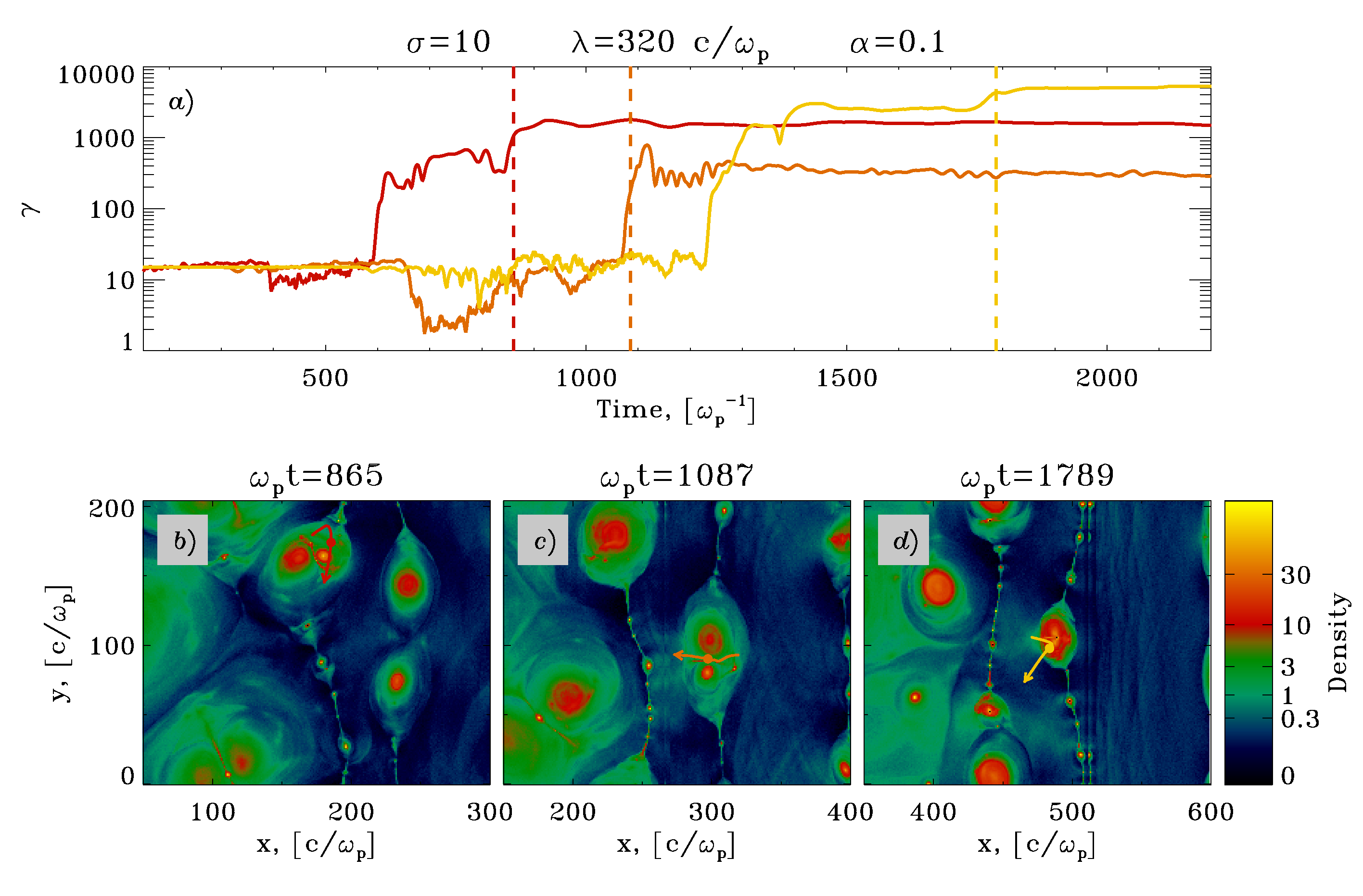}
\caption{Energy evolution of three representative high-energy positrons, in a striped flow with $\lambda=320\comp$, $\sigma=10$, and $\alpha=0.1$. See the supplementary material for an associate movie (\textsf{acceleration\_special.mov}). Panel (a): Energy histories of the selected positrons. Panels (b)-(d): 2D plots of particle number density, at the time when the three positrons in panel (a) are experiencing fast acceleration, as indicated by the vertical dashed  lines in panel (a). The positron trajectory is plotted with the same color coding as in panel (a). Along the positron track, the filled circle indicates the particle location at the time of the corresponding 2D density structure, and the arrow shows the velocity direction.}
\label{fig:accelfew}
\end{center}
\end{figure}

Although we find that the reconnection electric field at X-points is the main agent of particle acceleration, we have verified, by tracing the orbits of a large number of particles extracted directly from the simulation, that a variety of other mechanisms can play a role in particle energization. In \fig{accelfew}, we show the energy evolution of three representative positrons (panel (a)), together with the 2D density structure at the time when a given positron is being rapidly accelerated (panel (b) corresponds to the vertical red  dashed line in panel (a), panel (c) to orange, panel (d) to yellow). We find that the initial energy gain seen in the red and yellow curves of \fig{accelfew}(a) (at $\ompt\simeq600$ and $\ompt\simeq1250$, respectively) is due to the reconnection electric field that the particles experience upon their first encounter with an X-point, as described above. In contrast, the acceleration seen in the orange curve at $\ompt\simeq1100$ is powered by the anti-reconnection electric field at the X-point located in between a pair of merging islands, as discussed by  \citet{jaroschek_04} and \citet{oka_10}. In fact, \fig{accelfew}(c) shows that  the orange trajectory at $\ompt\simeq1087$ is confined in between two coalescing islands. In summary, reconnection and anti-reconnection at X-points govern the first stages of particle acceleration.

The late-time energization seen in the red and yellow curves of \fig{accelfew}(a) (at $\ompt\simeq850$ and $\ompt\simeq1800$, respectively) is due to processes not primarily related to the reconnection electric field. The positron shown in \fig{accelfew}(b) is accelerated by a kind of first-order Fermi process within magnetic islands, as discussed by \citet{drake_06}. The selected particle gains energy by reflecting from the ends of contracting islands following coalescence (its trajectory is orbiting around two merging islands in \fig{accelfew}(b)). In contrast, the positron in \fig{accelfew}(d) is accelerated at $\ompt\simeq1800$ via the standard first-order Fermi acceleration at shocks \citep{blandford_ostriker_78,bell_78,blandford_eichler_87}, by bouncing off a magnetic island (located in the middle of \fig{accelfew}(d)) that is advected into the shock with the upstream flow.

\section{Dependence on the Wind Wavelength and Magnetization}\label{sec:cond}
In this section, we investigate how the physical conditions in the wind can affect the structure of the shock and the physics of magnetic reconnection, with particular focus on the particle spectrum downstream from the hydrodynamic shock. We explore the dependence of our results on the stripe wavelength and the wind magnetization, for the case with no guide fields. We show that complete dissipation of the alternating fields (and transfer of field energy to the particles) occurs in all cases, but the width of the downstream particle spectrum depends on the properties of the wind.

\begin{figure}[tbp]
\begin{center}
\includegraphics[width=0.7\textwidth]{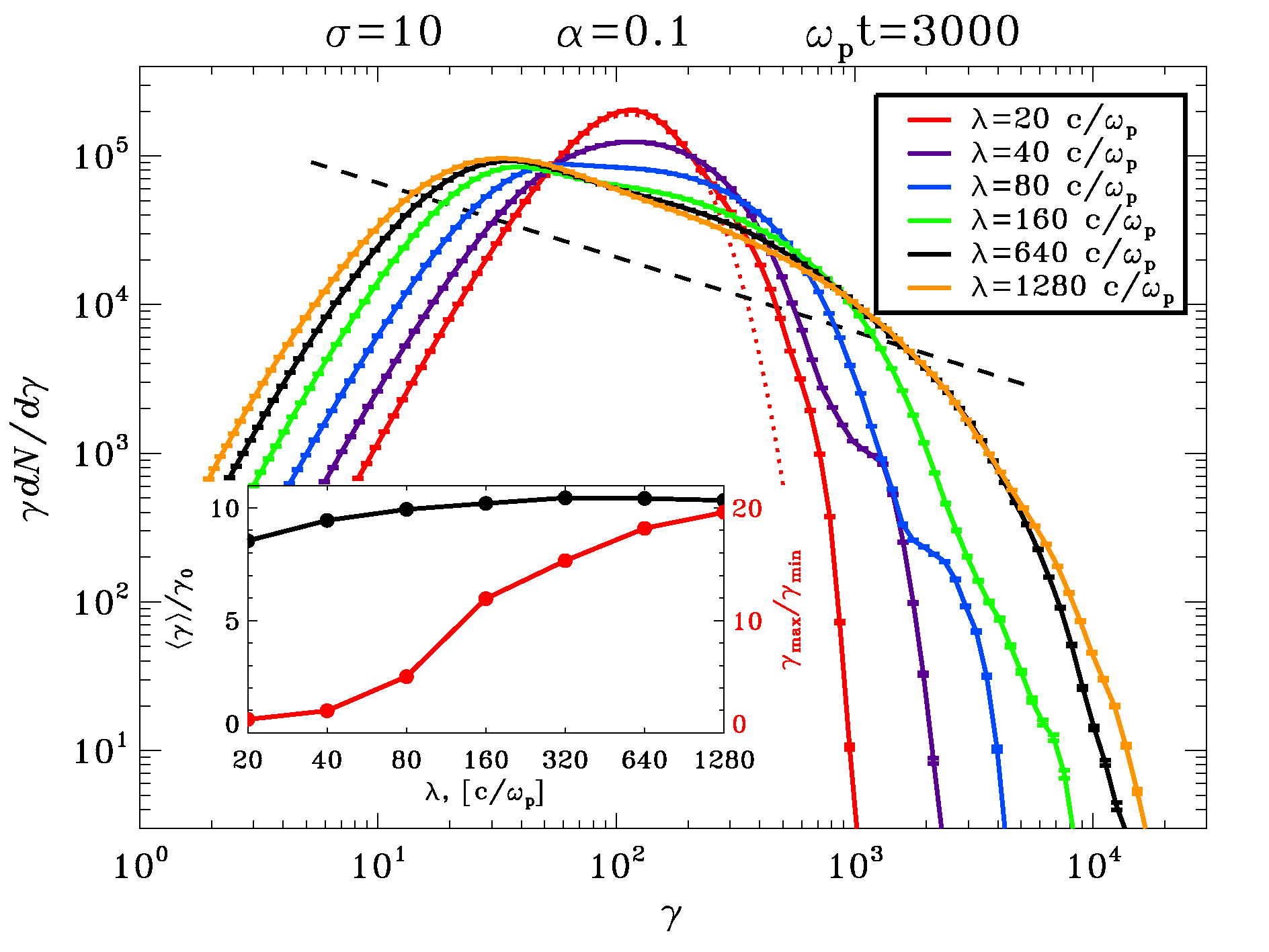}
\caption{Downstream particle spectrum for different values of $\lambda$, in a flow with $\sigma=10$ and $\alpha=0.1$. The dotted  red line is a Maxwellian with the same average energy as the spectrum colored in red (which refers to $\lambda=20\comp$); the dashed black line is a power law  with slope $p=1.5$. In the subpanel, the black line shows the average Lorentz factor of downstream particles $\langle\gamma\rangle$ as a function of $\lambda$ (axis on the left), whereas the red line presents the dependence of $\gamma_{\rm max}/\gamma_{\rm min}$ on $\lambda$ (axis on the right). Here, $\gamma_{\rm min}$ is the location where $\gamma dN/d\gamma$ peaks (i.e., where most of the particles reside), whereas  $\gamma_{\rm max}$ is the Lorentz factor where $\gamma^2 dN/d\gamma$ peaks (i.e., where most of the energy lies).}
\label{fig:speclam}
\end{center}
\end{figure}

\fig{speclam} shows how the post-shock energy spectrum changes for different stripe wavelengths $\lambda$, keeping fixed the magnetization $\sigma=10$ and the stripe-averaged field $\langle B_y\rangle_\lambda/B_0\simeq0.05$ (corresponding to $\alpha=0.1$). In all cases, the value of the post-shock mean particle Lorentz factor $\langle\gamma\rangle$ measured in our simulations (black line in the subpanel of \fig{speclam}) is consistent with full dissipation of the alternating fields, yielding $\langle\gamma\rangle\simeq\gamma_0\sigma$. Even though the mean kinetic energy per particle does not appreciably vary with $\lambda$, the shape of the spectrum does change, with a clear tendency for broader spectra at longer stripe wavelengths (as indicated also by the ratio $\gamma_{\rm max}/\gamma_{\rm min}$ between upper and lower spectral cutoffs, red line in the subpanel of \fig{speclam}). 

This trend can be easily understood by considering the structure of the flow, just upstream of the hydrodynamic shock. Here, the size of reconnection islands is constrained by the distance between neighboring current sheets, which is proportional to $\lambda$. Shorter wavelengths will then result in more numerous islands of smaller size, whereas fewer but bigger islands will be present for longer $\lambda$. Since an X-point exists in between each pair of neighboring islands (belonging to the same current sheet), the number of X-points per unit length (along the current sheet) will be larger for smaller $\lambda$. For short wavelengths, most of the incoming particles will likely pass in the vicinity of one of the numerous X-points, thus gaining energy from the reconnection electric field. All the particles will share similar energy evolutions, which results in a narrow Maxwellian-like distribution. On the other hand, for long wavelengths the energy evolution of different particles can be extremely diverse (as discussed in \S\ref{sec:accel}), depending on how far they pass from the closest X-point. Most of the particles stay far from X-points and remain cold, but the particles that interact with an X-point are accelerated to high energies by the reconnection electric field. This results in a broad energy spectrum.

\begin{figure}[tbp]
\begin{center}
\includegraphics[width=0.7\textwidth]{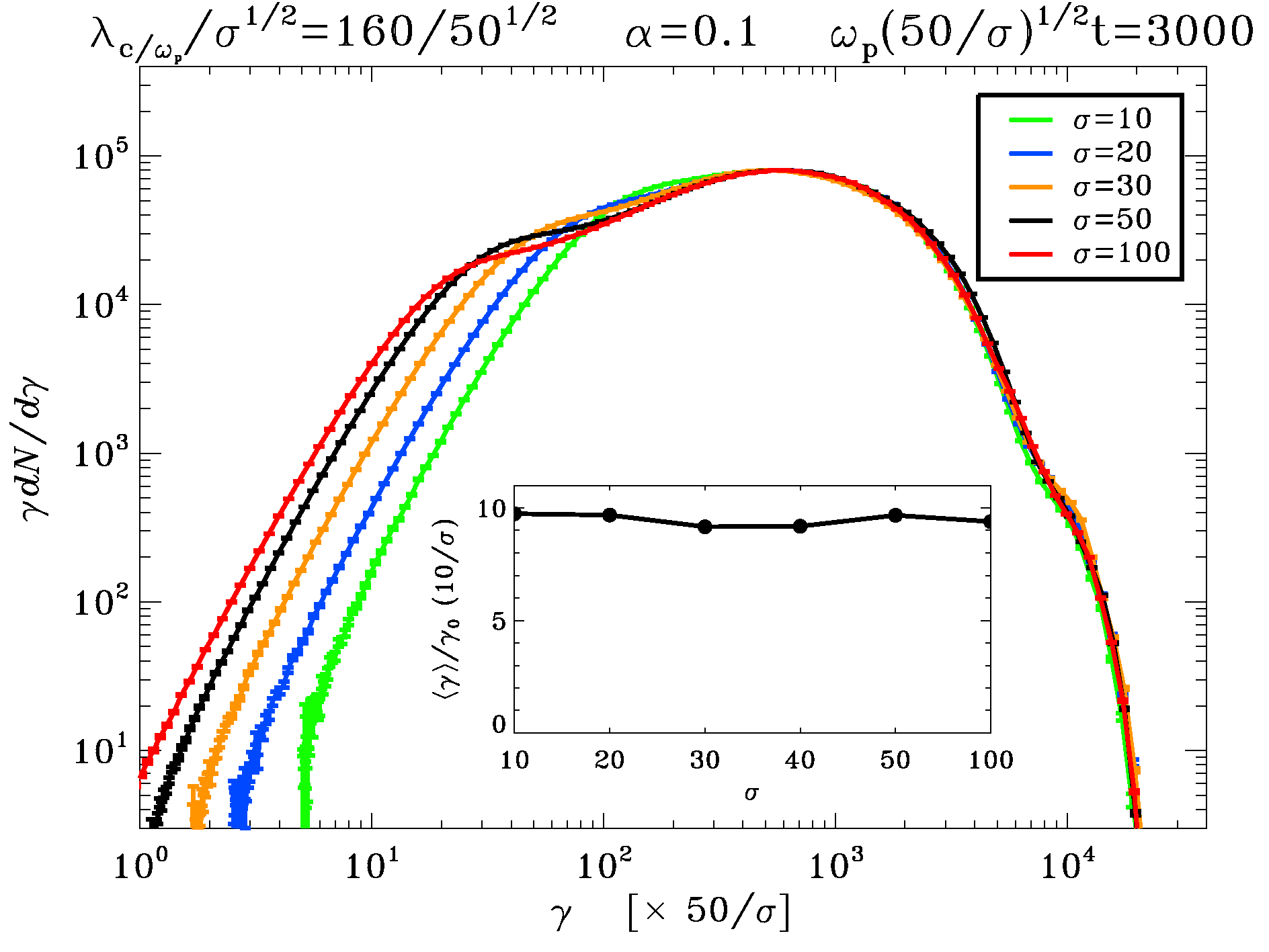}
\caption{Downstream particle spectrum for different values of $\lambda$ and $\sigma$, but such that the ratio $\lambda/(r_L\sigma)\simeq22.5$ is kept fixed (as well as $\alpha=0.1$). Here, $\lambda/(r_L\sigma)$ is roughly the wind wavelength in units of the \tit{post-shock} plasma skin depth. The comparison is performed at the same time, in units of the relativistic plasma frequency of the  \tit{post-shock}  flow ($\simeq\omega_{\rm p}/\sqrt{\sigma}$). Spectra are shifted along the $x$-axis by $50/\sigma$ to facilitate comparison with the case $\sigma=50$. The black line in the subpanel shows the average downstream Lorentz factor $\langle\gamma\rangle$ as a function of $\sigma$.}
\label{fig:scalesig}
\end{center}
\end{figure}

We find that  the threshold between short and long wavelengths depends on the stripe wavelength and the wind magnetization via the combination $\lambda/(r_L\sigma)$, namely the stripe wavelength measured in units of the  \tit{post-shock} plasma skin depth. In \fig{scalesig} we confirm that the main properties of the shock, and in particular the high-energy end of the downstream particle spectrum, are relatively insensitive to variations in $\lambda$ or $\sigma$, provided that the ratio $\lambda/(r_L\sigma)$ is kept constant. When varying $\lambda/(r_L\sigma)$, a Maxwellian-like spectrum is obtained for $\lambda/(r_L\sigma)\lesssim$ a few tens, whereas in the limit $\lambda/(r_L\sigma)\gg1$ the spectrum approaches a broad power-law tail of index $1<p<2$, extending from $\gamma_{\rm min}\simeq\gamma_0$ up to $\gamma_{\rm max}\simeq\gamma_0\sigma^{1/(2-p)}$.

The minor bump emerging at high energies in the spectra of Figs.~\fign{speclam} and \fign{scalesig} is populated by particles that are accelerated via shock-drift acceleration \citeaffixed{chen_75,webb_83,begelman_kirk_90}{SDA;} at the hydrodynamic shock. They gain energy from the stripe-averaged motional electric field $\langle  E_z\rangle_\lambda\simeq\langle  B_y\rangle_\lambda$, while gyrating around the shock, before being advected downstream by the stripe-averaged magnetic field. Given the limited number of acceleration cycles, the SDA component does not extend in time to higher energies. So, SDA is not a promising candidate to produce broad power-law tails.

\section{Three-Dimensional Simulations}\label{app:3d}
The results presented so far are based on 2D experiments with magnetic field initialized in the plane of the simulations. This is the best configuration to study the tearing-mode instability, whose wavevector lies in the plane of the field, but it artificially inhibits the growth of cross-plane instabilities, which may also be important for the structure of the shock. In this section, we study the 3D physics of shock-driven reconnection, with particular emphasis on particle acceleration. The simulation setup parallels very closely what we described in \S\ref{sec:setup}, with the alternating field oriented initially along the $y$ direction and no guide field. We adopt our fiducial values for $\sigma=10$ and $\alpha=0.1$, and we investigate $\lambda=80\comp$ and $\lambda=160\comp$. For each value of the stripe wavelength, the extent of the simulation domain along $y$ and $z$ is chosen such that our results are not artificially affected by the periodicity of our boundary conditions (see Appendix B).

\begin{figure*}[tbp]
\begin{center}
\includegraphics[width=0.95\textwidth]{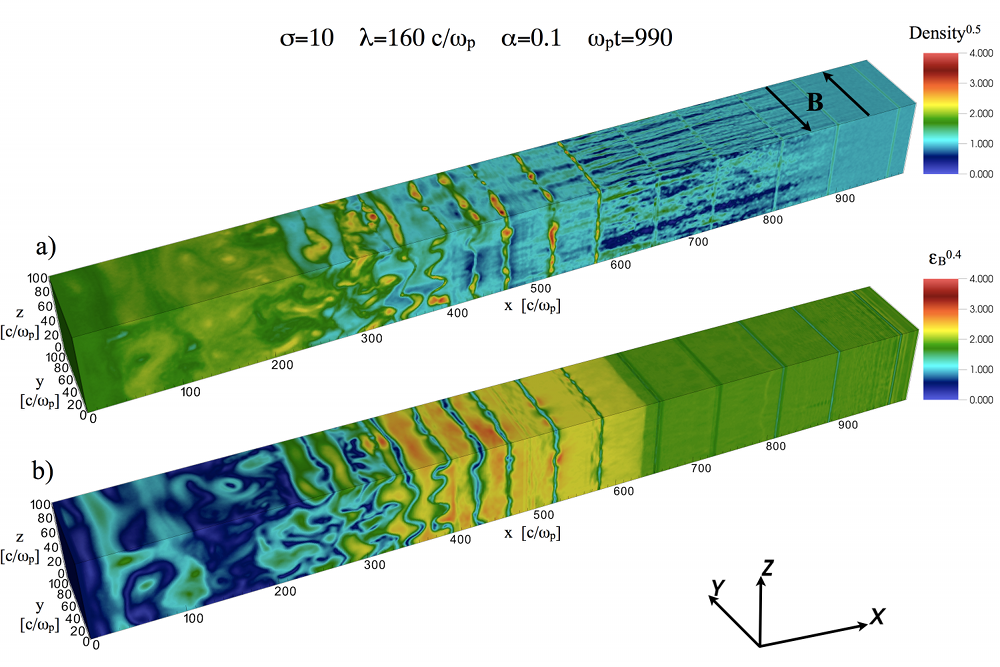}
\caption{Internal structure of the flow at $\ompt=990$, from the 3D simulation of a striped wind with $\lambda=160\comp$, $\sigma=10$, and $\alpha=0.1$. See the supplementary material for associate movies (\textsf{3d\_density.mov} and \textsf{3d\_magnetic\_energy.mov}) that follow the complete evolution of the shock. The magnetic field is initialized in the $xy$ plane, as shown by the black arrows in panel (a). The hydrodynamic shock is located at $x\simeq250\comp$, and the fast MHD shock at $x\simeq630\comp$. We show the 3D plots of: (a) particle number density, in units of the upstream value, with color scale stretched to enhance contrast; (b) magnetic energy fraction $\epsilon_B\equiv B^2/8 \pi \gamma_0 m n_{c0} c^2$, with color scale stretched to enhance contrast.}
\label{fig:fluid3d}
\end{center}
\end{figure*}

\fig{fluid3d} shows the structure of the shock at $\ompt=990$, for a striped wind of wavelength $\lambda=160\comp$. In agreement with the results presented in \S\ref{sec:shock} for 2D simulations, the structure of the flow in \fig{fluid3d} shows the presence of two shocks, a fast MHD shock located at $x\simeq630\comp$ and a hydrodynamic shock at $x\simeq250\comp$. The passage of the fast shock through the incoming current sheets initiates magnetic field annihilation. The dissipation of alternating fields proceeds as the flow propagates from the fast to the hydrodynamic shock, and little magnetic energy remains downstream from the hydrodynamic shock (see \fig{fluid3d}(b) at $x\lesssim250\comp$), in agreement with our 2D results. 

As the flow recedes behind the fast shock, the structure of the current sheets is affected by the growth of two competing modes. In the $xy$ plane of the magnetic field, the tearing-mode instability breaks the current sheet into a sequence of high-density islands (see \fig{fluid3d}(a)), separated by X-points where the field lines tear and reconnect. In the $xz$ plane orthogonal to the field, the folding of the current sheet seen in \fig{fluid3d} is governed by the drift-kink instability, driven by the current of fast-drifting plasmas in a thin sheet \citep{daughton_98,zenitani_05,zenitani_07}. The 2D geometry employed in the previous sections was chosen to select the tearing mode, and suppress the drift-kink mode. 

The nature of the instability that dominates the process of field annihilation should leave an imprint on the resulting particle distribution. \citet{zenitani_07} have shown with PIC simulations  that  field dissipation due to the drift-kink instability does not result in nonthermal particle acceleration, the plasma just being heated. This is because the field lines remain straight, so that all particles gain energy at the same rate. Nonthermal particles are produced only by the tearing mode \cite{zenitani_01}. In our 2D simulations with in-plane fields, i.e., the geometry required to capture the tearing mode, we have shown that particles are accelerated to nonthermal energies by the reconnection electric field. They initially move along the $z$ direction parallel to the reconnection field, and then they drift along $y$ from a given X-point into the closest magnetic island. In 3D, one could argue that the folding of the current sheet introduced by the drift-kink instability in the $xz$ plane (not resolved in 2D) may deflect the accelerating particles out of the current sheet, thus suppressing their energization. 

\begin{figure}[tbp]
\begin{center}
\includegraphics[width=0.7\textwidth]{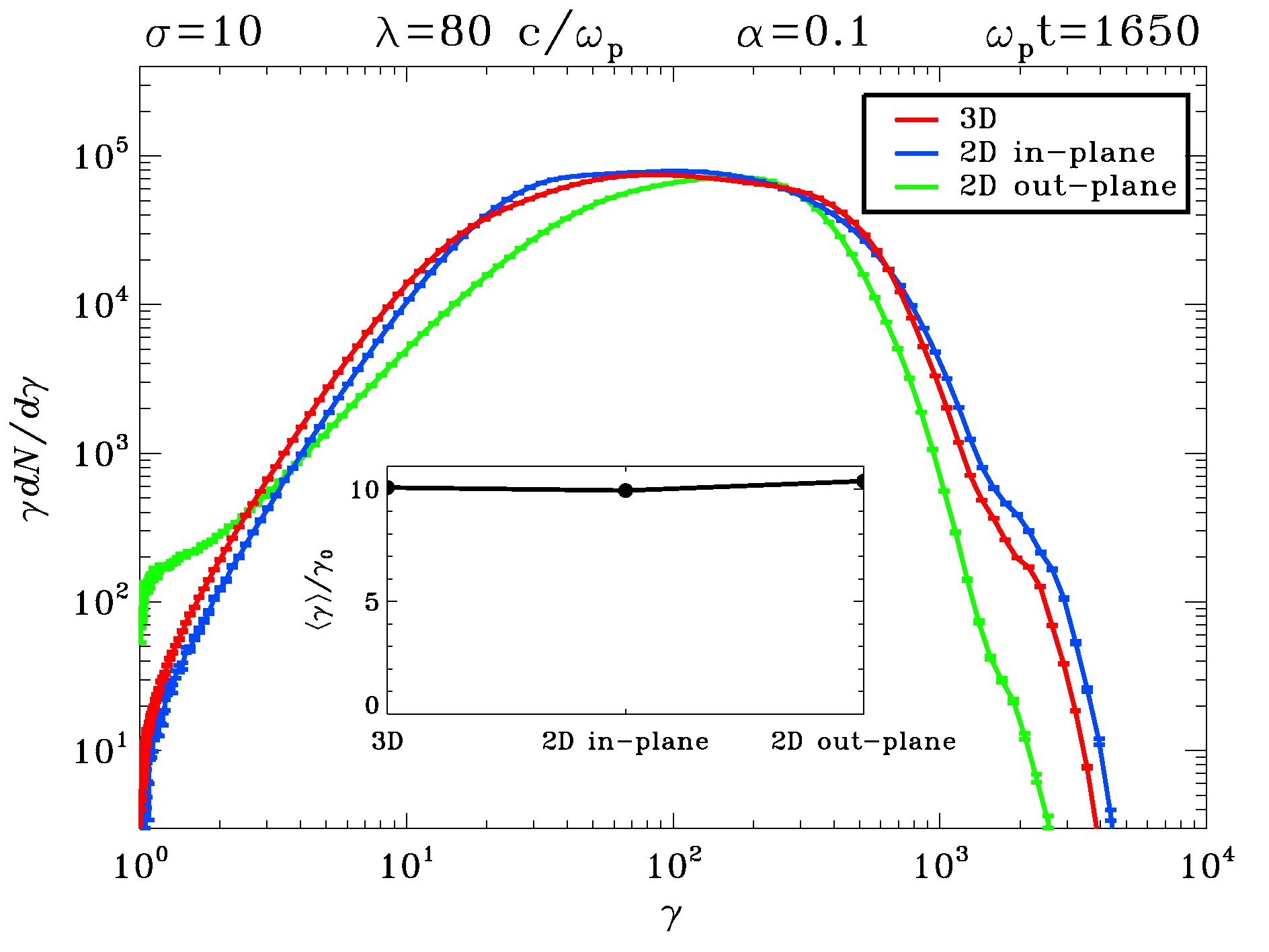}
\caption{In a striped wind with $\lambda=80\comp$, $\sigma=10$, and $\alpha=0.1$, comparison of the downstream particle spectrum for different magnetic field configurations: 3D simulation (red) and 2D simulations with either in-plane (blue) or out-of-plane (green) magnetic fields. The black line in the subpanel shows the average downstream Lorentz factor  $\langle\gamma\rangle$ for the different cases.}
\label{fig:spec3d}
\end{center}
\end{figure}

In \fig{spec3d}, we address this important issue, by comparing for $\lambda=80\comp$ the post-shock particle spectrum of a 3D simulation (red line) to the results of 2D experiments with in-plane or out-of-plane magnetic fields (blue and green lines, respectively). The excellent agreement between 3D and 2D in-plane results (red and blue line, respectively) suggests that the physics of particle acceleration by shock-driven reconnection is captured extremely well by our 2D simulations with in-plane fields.\footnote[3]{The agreement between red and blue lines in the high-energy bump at $\gamma\gtrsim2000$ also suggests that the physics of SDA is correctly described by our 2D simulations with in-plane fields.} In turn, this implies that, at least for $\lambda=80\comp$, the maximum energy of accelerated particles is constrained by the distance between X-points and islands (in the $xy$ plane), rather than by the current sheet folding in the $xz$ plane.

Finally, as we discuss in \S\ref{sec:guide}, the tension force of a guide field (along $z$, in our geometry) can easily stabilize the drift-kink mode, with little or no effect on the tearing mode \citep{zenitani_08}. In the presence of a guide field, the 3D physics of shock-driven reconnection should be described very accurately by 2D simulations with alternating fields lying in the simulation plane. 

\section{Dependence on the Guide Field Strength}\label{sec:guide}
The results presented so far are obtained in the absence of a guide field, i.e., the component of  magnetic field aligned with the particle current in the sheets. In this section, we investigate the effects of a guide field component  on the structure of the termination shock. A guide field  is not usually expected in the context of pulsar winds \citeaffixed{petri_kirk_05}{but see}, but it may be important in gamma-ray burst and blazar jets, where it could be even stronger than the alternating component. We assume that the alternating fields are oriented either along or perpendicular to the $xy$ simulation plane (so that the guide field is along $z$ or $y$, respectively). In addition to the alternating fields  of strength $B_0$, we  initialize in the pre-shock flow a uniform guide field of  intensity $B_g$ (with $0.0\leq B_g/B_0 \leq 1.0$), both within and outside the current sheets. 

The role of the guide field in the case of \textit{undriven reconnection} has been investigated by \citet{zenitani_08} by means of linear analysis and PIC simulations. They found that the tearing mode is relatively insensitive to the strength of the guide field, whereas the relativistic drift-kink instability is stabilized by the tension force of the guide field if $B_g/B_0\gtrsim0.3$. In this section, we perform a similar analysis for \textit{shock-driven reconnection} at the termination shock of striped winds,  focusing on the process of particle acceleration. For alternating fields lying in the plane of the simulations, the drift-kink mode is not resolved, and we can isolate the effects of the guide field on the tearing mode. Instead, if the alternating fields are orthogonal to the computational domain, we can study whether the guide field can stabilize the drift-kink mode.

\begin{figure}[tbp]
\begin{center}
\includegraphics[width=\textwidth]{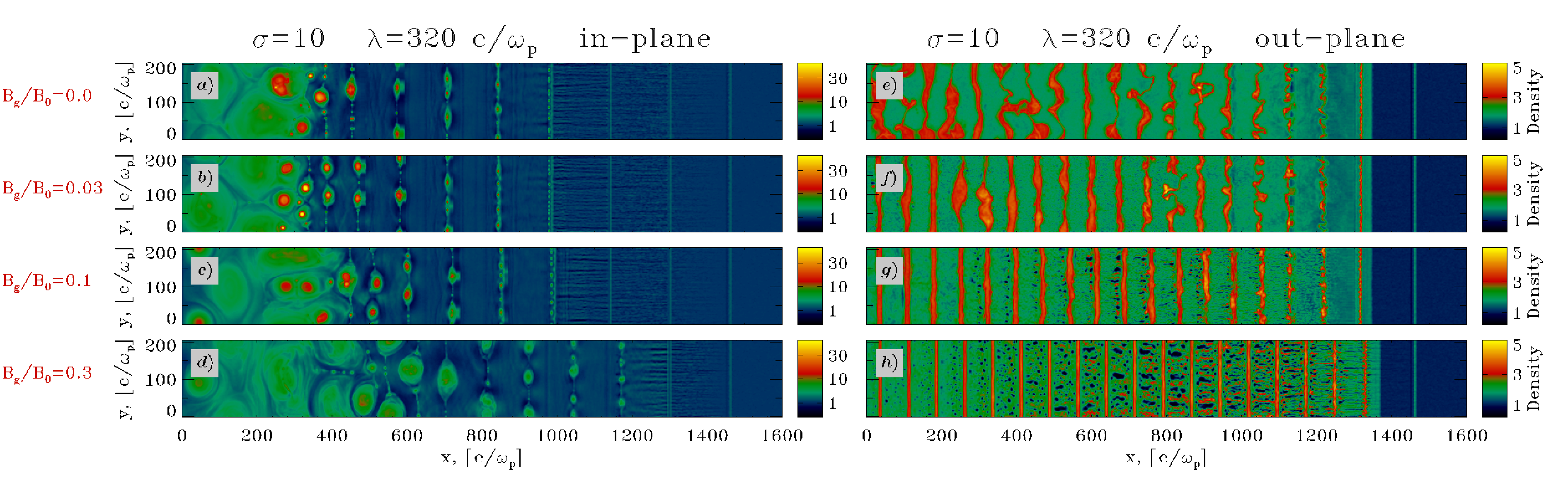}
\caption{2D structure of the particle number density (in units of the upstream value), for alternating fields lying either in the simulation plane (left column) or orthogonal to the plane (right column). We vary the guide field strength $B_g/B_0$ as indicated on the left. The figure refers to a flow with $\lambda=320\comp$, $\sigma=10$, and $\alpha=0.1$, at time $\ompt=1500$.}
\label{fig:fluidguide} 
\end{center}
\end{figure}

\fig{fluidguide} shows the structure of the flow in the two cases (left column for in-plane alternating fields, right column for out-of-plane fields), for different strengths of the guide field, ranging from $B_g/B_0=0.0$ up to $0.3$. The wavelength and magnetization of the striped wind are $\lambda=320\comp$ and $\sigma=10$  in all cases. The left column shows that, in the case of in-plane alternating fields (so that the guide field is along $z$), the development of shock-driven reconnection and the resulting structure of the flow are nearly insensitive to the strength of the guide field. For all the values of $B_g$ we investigate ($B_g/B_0$ from 0.0 to 0.3, from top to bottom), a fast shock propagates into the striped wind, compressing the incoming current sheets and triggering the tearing mode, which results in the magnetic islands seen in the density structure of \fig{fluidguide} (left column). When the size of magnetic islands becomes comparable to the distance between neighboring current sheets, the striped structure of the flow is erased, and the main shock forms (at $x\simeq400\comp$ in the left column of \fig{fluidguide}), as described in \S\ref{sec:shock}. The alternating fields annihilate completely at the shock, whereas the guide field  is compressed, in agreement with MHD jump conditions. For stronger guide fields, the magnetic pressure in the post-shock plasma is larger, and the shock propagates at a higher speed. This explains why the location of the main shock shifts toward $+\bmath{\hat{x}}$ as the guide field strength increases from 0.0 to 0.3 (panel (a) through (d) in the left column). 

\begin{figure}[tbp]
\begin{center}
\includegraphics[width=0.7\textwidth]{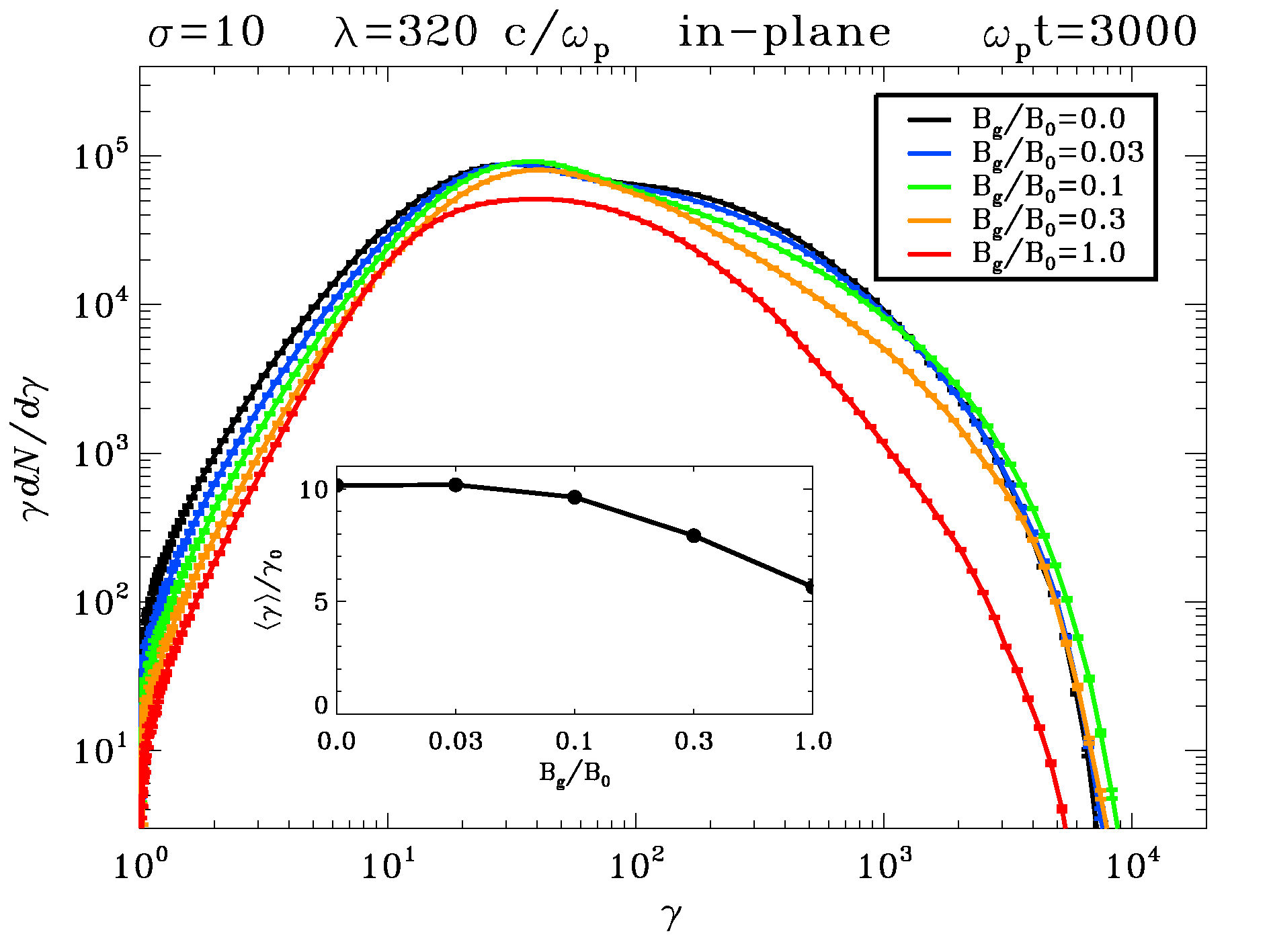}
\caption{Downstream particle spectrum at $\ompt=3000$ for different values of the guide field strength $B_g/B_0$, in a flow with $\lambda=320\comp$, $\sigma=10$,  $\alpha=0.1$, and alternating fields lying within the simulation plane. In the subpanel, the black line shows the mean Lorentz factor $\langle\gamma\rangle/\gamma_0$ of post-shock particles as a function of $B_g/B_0$.}
\label{fig:specy}
\end{center}
\end{figure}

In summary, the overall flow structure in the case of in-plane alternating fields -- the configuration required to capture  the tearing mode instability -- is not appreciably changed by the presence of the guide field, up to $B_g/B_0=0.3$. This is confirmed by  \fig{specy}, where we compare the post-shock particle energy spectra for values of $B_g/B_0$ ranging from 0.0 to 1.0. We find that the particle spectrum departs from the result obtained in the absence of a guide field (black line) only for $B_g/B_0\gtrsim0.3$ (yellow line for $B_g/B_0=0.3$, red line for $B_g/B_0=1.0$). For strong guide fields the rate of particle inflow into the X-points is smaller, since the plasma should be moving across the guide field lines. This results in fewer particles being accelerated by the reconnection electric field. This explains why the high-energy component  in the particle spectra of \fig{specy} is suppressed for $B_g/B_0\gtrsim0.3$. In turn, this corresponds to the decrease of  mean particle Lorentz factor  shown in the subpanel of \fig{specy}, with respect to the value $\langle\gamma\rangle\simeq\gamma_0(\sigma+1)$ expected in the case of complete field annihilation.\footnote[4]{The different normalization of the spectral peak for $B_g/B_0=1.0$ (red line in \fig{specy}) results from the fact that the post-shock flow is more magnetized, for extreme values of the guide field. It follows that the shock speed is higher, so the post-shock density is lower.}

\begin{figure}[tbp]
\begin{center}
\includegraphics[width=0.7\textwidth]{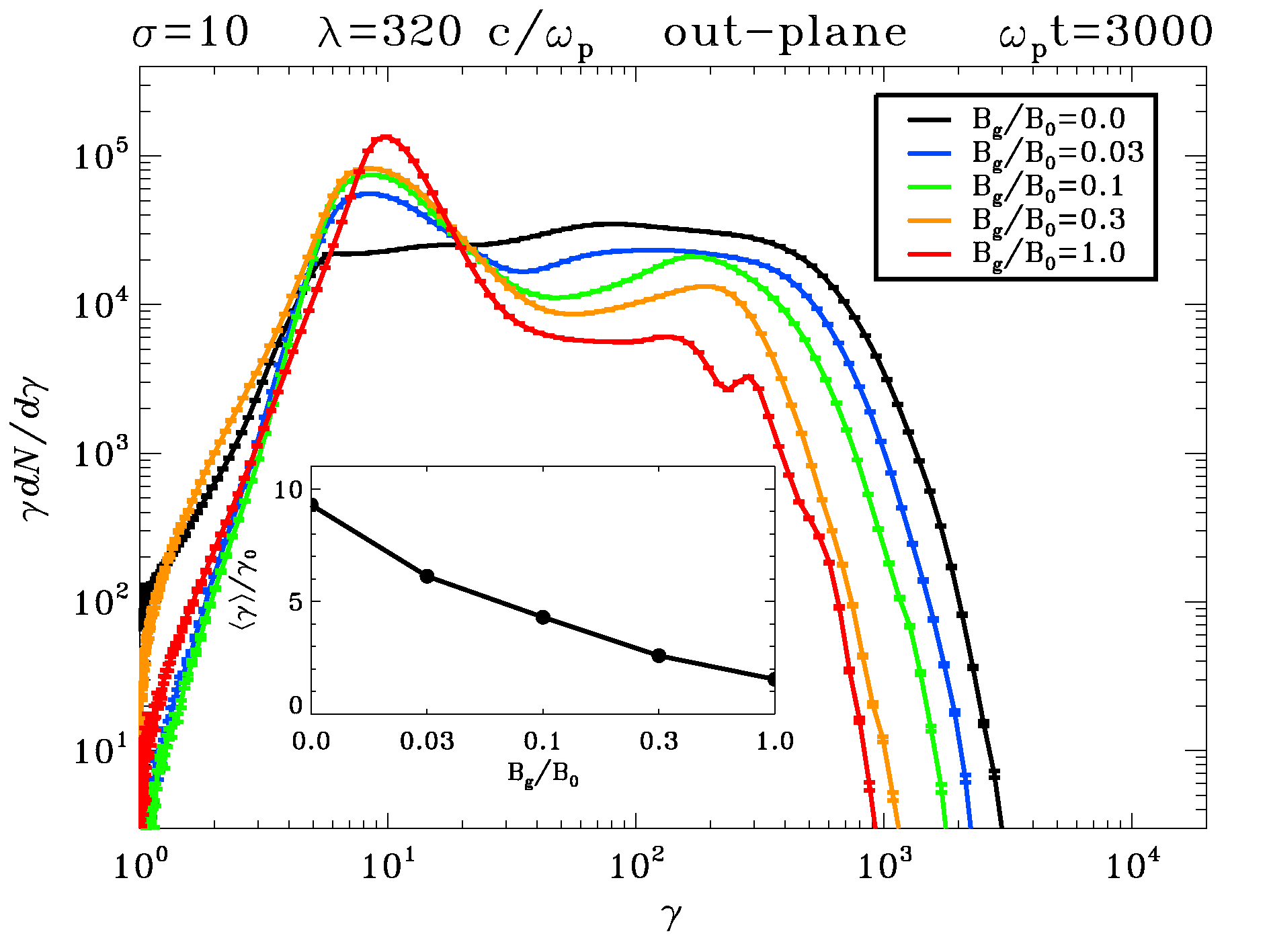}
\caption{Downstream particle spectrum at $\ompt=3000$ for different values of the guide field strength $B_g/B_0$, in a flow with $\lambda=320\comp$, $\sigma=10$,  $\alpha=0.1$, and alternating fields orthogonal to the simulation plane. In the subpanel, the black line shows the mean Lorentz factor $\langle\gamma\rangle/\gamma_0$ of post-shock particles as a function of $B_g/B_0$.}
\label{fig:specz}
\end{center}
\end{figure}

So far we have discussed the effects of a guide field component on the development of the tearing instability, the only mode expected for alternating fields lying in the simulation plane. The right column of \fig{fluidguide} shows instead the effect of the guide field in the case of alternating fields oriented along $z$, i.e., perpendicular to the simulation plane. For this geometry, the guide field is oriented along $y$. The flow structure in the absence of a guide field ($B_g/B_0=0.0$ in panel (e) of \fig{fluidguide}) shows a fast MHD shock located at $x\simeq1400\comp$, which compresses the incoming flow and triggers the drift-kink mode,  manifested in the folding of the current sheets downstream from the shock.\footnote[5]{The speed of the shock is much larger here than in the case of in-plane fields with $B_g/B_0=0.0$ of \fig{fluidguide}(a). This results primarily from the different adiabatic index imposed by the 2D geometry of our simulations. If the alternating fields are in the simulation plane, the particle gyro-motion around the field will involve the third dimension (i.e., orthogonal to the plane), so that the post-shock plasma will be isotropic in 3D. In contrast, for out-of-plane alternating fields, the motion of the plasma in the presence of the field is confined to the 2D plane of the simulations, resulting in a 2D adiabatic index. The fact that the tearing mode is more effective than the drift-kink mode in converting field energy to particle energy further contributes to the difference in shock speed between panel (a) and (e).} The right column of \fig{fluidguide} clearly shows that the drift-kink mode is stabilized once the guide field exceeds $B_g/B_0\simeq0.1$, in agreement with the conclusions by \citet{zenitani_08}. For $B_g/B_0=0.3$ in \fig{fluidguide}(h), the current sheets downstream from the shock do not show any sign of the kink mode. Due to the suppression of the drift-kink mode for $B_g/B_0\gtrsim0.1$, most of the post-shock energy stays in the magnetic field, and the plasma remains cold. This is clearly shown in the particle energy spectra of \fig{specz}, for different strengths of the guide field. As the value of $B_g/B_0$ increases, the spectrum passes from a flat distribution with mean Lorentz factor $\simeq\gamma_0(\sigma+1)$ (black line for $B_g/B_0=0.0$), as expected for efficient field annihilation, to a mono-energetic distribution peaking at $\gamma\simeq\gamma_0$ (red line for $B_g/B_0=1.0$), as in the case of negligible field dissipation. The subpanel of \fig{specz} shows that the suppression of the drift-kink mode, and the resulting inhibition of particle heating, occurs for guide field strengths as small as $B_g/B_0\simeq0.03$.

\section{Summary and Discussion}\label{sec:disc}
We have explored by means of 2D and 3D PIC simulations the internal structure and acceleration properties of relativistic shocks that propagate in an electron-positron striped wind, i.e., a flow consisting of stripes of alternating field polarity separated by current sheets of hot plasma. 

We find that a fast MHD shock propagates into the pre-shock striped flow, compressing the incoming current sheets and initiating the process of \textit{driven magnetic reconnection}, via the tearing-mode instability. Reconnection islands seeded by the passage of the fast shock grow and coalesce, while magnetic energy is dissipated at X-points located in between each pair of islands. When reconnection islands grow so big to occupy the entire region between neighboring current sheets, the striped structure of the flow is erased, and a hydrodynamic shock forms. By this point, the energy stored in the alternating fields has been entirely transferred to the particles via shock-driven magnetic reconnection, so the post-shock fluid behaves like an unmagnetized plasma. In our 2D and 3D simulations, we find that complete annihilation of the alternating fields occurs irrespective of the stripe wavelength $\lambda$ or the magnetization $\sigma$.

The main agent of particle energization is the reconnection electric field, as particles drift from a given X-point into the closest island. Whether all particles have comparable energy gains, or only a few particles are accelerated to high energies, and the majority stay cold, depends sensitively on the properties of the wind.  In the absence of guide fields, we find that the shape of the post-shock spectrum depends primarily on the combination $\lambda/(r_L\sigma)$, where $r_L$ is the relativistic Larmor radius in the striped wind. For small values of $\lambda/(r_L\sigma)$ ($\lesssim$ a few tens), the spectrum resembles a Maxwellian distribution with mean energy $\langle\gamma\rangle\simeq\gamma_0\sigma$, where $\gamma_0$ is the bulk Lorentz factor of the pre-shock flow. In the limit of very large values of $\lambda/(r_L\sigma)$ ($\gtrsim$ a few hundreds), the spectrum approaches a broad power-law tail with slope $p\simeq1.5$, extending  from  $\gamma_{\rm min}\simeq\gamma_0$ up to $\gamma_{\rm max}\simeq \gamma_0 \sigma^{1/(2-p)}$. Our results are nearly insensitive to the strength of the guide field $B_g$, in the regime $B_g/B_0\lesssim0.3$, where $B_0$ is the magnitude of the alternating component. For strong guide fields (i.e., $B_g/B_0\gtrsim0.3$) the rate of particle inflow into the X-points is smaller, since the plasma should be moving across the guide field lines. This results in fewer particles being accelerated by the reconnection electric field, and the post-shock spectrum shifts to lower energies, everything else being fixed.

The particle energy spectra extracted from our simulations can be used directly to interpret the radiative signature of astrophysical nonthermal sources, and in particular of PWNe. The radio spectrum of the Crab Nebula, the prototype of PWNe, requires a population of nonthermal particles with a flat spectral slope ($p\simeq1.5$), extending from $10^2\unit{MeV}$ up to $10^5\unit{MeV}$. Our results suggest that, for $p\simeq1.5$, the particle spectrum will extend over three decades in energy if the wind magnetization at the termination shock is $\sigma\gtrsim30$. In addition, broad particle spectra are generated only if  $\lambda/(r_L\sigma)\gtrsim$ a few tens. At the termination shock of the pulsar wind ($R=R_{TS}$) we have
\be\label{eq:wind}
\frac{\lambda}{r_L \sigma}\simeq4\pi\kappa \frac{R_{LC}}{R_{TS}}~~,
\ee
where $R_{LC}=c/\Omega$ is the light cylinder radius ($\Omega$ is the pulsar rotational frequency), and $\kappa$ is the so-called multiplicity in the wind (i.e., the ratio of the actual density to the \possessivecite{goldreich_julian_69} density). For the Crab, $R_{TS}\simeq5\times10^8R_{LC}$ \cite{hester_02}, and most available models estimate $\kappa\simeq10^4-10^6$ \citep{bucciantini_11}. Based on our findings,  the resulting value of $\lambda/(r _L \sigma)\lesssim0.01$ would yield a Maxwellian-like spectrum, at odds with the wide flat spectrum required by observations. If radio-emitting electrons are produced in the equatorial plane by shock-driven reconnection, a revision of the existing estimates of $\kappa$ is required. In this respect, we point out that the values of $\kappa$ quoted above are averages over latitude, and one cannot exclude that particle injection into the pulsar wind is highly anisotropic, with multiplicity as large as $10^8$ along the equatorial plane. This is not in conflict with the observed lack of gradients in the radio spectral slope of the Crab \cite{bietenholz_kronberg_92}, since strong fluid motions downstream from the termination shock would quickly fill the entire nebula with the long-lived radio-emitting electrons \cite{delzanna_04}. Alternatively, the radio part of the spectrum may be produced at higher latitudes, where the termination shock is closer to the pulsar, and the ratio in \eq{wind} becomes larger.  In summary, our findings place important constraints on the magnetospheric physics of pulsars, and they provide physically-grounded inputs for models of nonthermal emission in PWNe.
\ack
L.S. is supported by NASA through Einstein
Postdoctoral Fellowship grant number PF1-120090 awarded by the Chandra
X-ray Center, which is operated by the Smithsonian Astrophysical
Observatory for NASA under contract NAS8-03060. A.S. is supported by NSF
grant AST-0807381 and NASA grant NNX10AI19G. The simulations presented in this
article were performed on computational resources supported
by the PICSciE-OIT High Performance Computing
Center and Visualization Laboratory, and at  National
Energy Research Scientific Computing Center, which is
supported by the Office of Science of the US Department
of Energy under contract No. DE-AC02-05CH11231.

\appendix
\vspace{-0.1in}
\section{Dependence on the Current Sheet Thickness}\label{app:thick}
\begin{figure}[htbp]
\begin{center}
\includegraphics[width=0.7\textwidth]{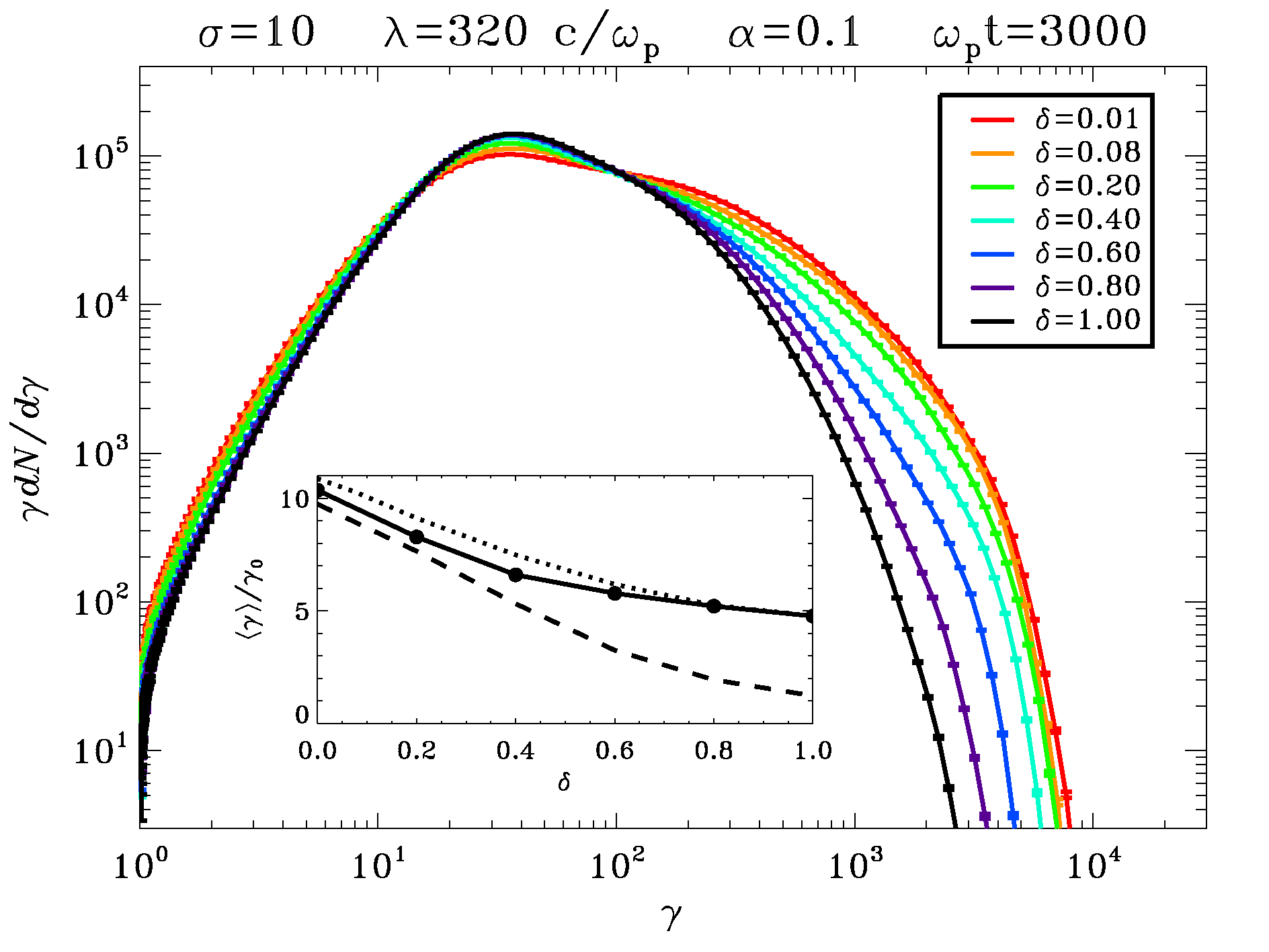}
\caption{Downstream particle spectrum at $\ompt=3000$ for different values of the current sheet width $\Delta$, parameterized in terms of the fractional thickness $\delta\equiv\pi \Delta/\lambda$, in a flow with $\lambda=320\comp$, $\sigma=10$, and $\alpha=0.1$. We vary $\delta$ from $\delta=0.01$ (corresponding to $\Delta\simeq\comp$) up to $\delta=1.0$ (corresponding to $\Delta\simeq\lambda/\pi$, i.e., comparable to the half-wavelength of the stripes). The solid  line in the subpanel shows the average downstream Lorentz factor  $\langle\gamma\rangle$, i.e., the  post-shock mean \tit{kinetic} energy per particle. The dotted line shows the  pre-shock mean \tit{kinetic+electromagnetic} energy per particle, and the dashed line is the  pre-shock mean \tit{electromagnetic} energy per particle. The different spectra are normalized such that to contain the same number of particles.}
\label{fig:specthick}
\end{center}
\end{figure}

The results presented in the main body of the paper have been obtained assuming that the thickness of  current sheets in the wind is comparable to the relativistic skin depth, i.e., $\Delta\simeq\comp$. This  value is large enough so that the process of undriven reconnection does not significantly modify the flow structure ahead of the termination shock. On the other hand, since $\delta\equiv \pi\Delta/\lambda\ll1$, the current sheets only occupy a small fraction of the striped flow, and most of the pre-shock particles and energy are contributed by the material in the cold wind. In \fig{specthick}, we study the dependence of the post-shock particle spectrum on the current sheet thickness, by increasing the fractional width $\delta$ from 0.01 (as adopted so far for $\lambda=320\comp$) up to 1.0.

As shown in \fig{specthick}, for $\delta\lesssim0.2$ the shape of the post-shock spectrum is insensitive to the sheet thickness. In this case, both the evolution of the flow across the shock, and the physics of particle acceleration by reconnection, proceed in the same way as described in \S\ref{sec:shock} and \S\ref{sec:accel}. The only difference is in the size of magnetic islands just downstream from the fast MHD shock, since the wavelength of the most unstable tearing mode scales linearly with the sheet thickness \citep{zenitani_07}. 

As $\delta$ increases beyond 0.2, the high-energy end of the post-shock spectrum gets progressively depopulated, and it shifts to lower energies. At the same time, the post-shock mean \tit{kinetic} energy per particle (solid line in the subpanel) decreases.\footnote{Strictly speaking, the solid line also includes  the contribution of the rest-mass energy. In this section, we simply call ``kinetic energy'' (as opposed to ``electromagnetic energy'') the sum of the kinetic and rest-mass energy.} However, this is not the result of a smaller efficiency in converting electromagnetic energy to particle energy via reconnection. Rather, it comes from the fact that, with increasing $\delta$, the mean \tit{kinetic+electromagnetic} energy per particle in the pre-shock flow becomes smaller, everything else being fixed (dotted line in the subpanel). This is because in our setup the current sheets are overdense by a factor of four with respect to the cold wind, so that the mean particle energy within current sheets is smaller by the same amount. In the subpanel, the agreement between the solid line (post-shock mean \tit{kinetic} energy per particle) and the dotted line (pre-shock mean \tit{kinetic+electromagnetic} energy per particle) suggests that complete field dissipation occurs regardless of $\delta$. However, as $\delta$ increases, a smaller fraction of the pre-shock energy is in electromagnetic form (dashed line in the subpanel), which corresponds to a smaller effective magnetization of the upstream flow. It follows that the reconnection electric field will be weaker, and particles will not be accelerated to energies as high as in the case of thin current sheets.

\vspace{-0.1in}
\section{Dependence on the Transverse Size of the Simulation Box}\label{app:my}
\begin{figure}[htbp]
\begin{center}
\includegraphics[width=0.7\textwidth]{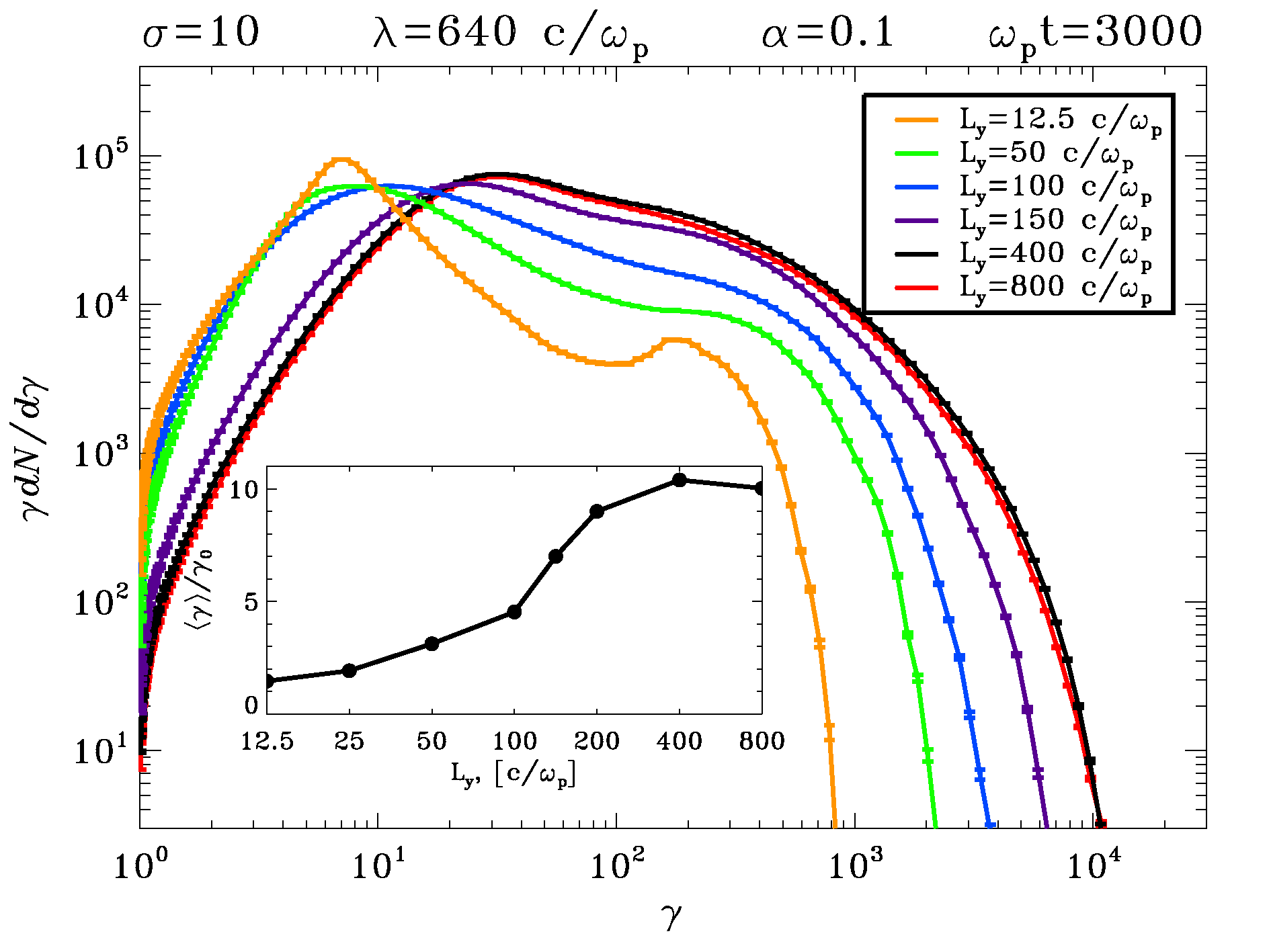}
\caption{Downstream particle spectrum at $\ompt=3000$ for different values of the transverse size $L_y$ of our simulation box, in a flow with $\lambda=640\comp$, $\sigma=10$, and $\alpha=0.1$. We vary $L_y$ from $L_y=800\comp$ down to $L_y=12.5\comp$, which approaches the 1D setup of \citet{petri_lyubarsky_07}. The black line in the subpanel shows the average downstream Lorentz factor  $\langle\gamma\rangle$ as a function of the box width $L_y$.}
\label{fig:specmy}
\end{center}
\end{figure}

In \fig{specmy} we investigate the dependence of our findings on the transverse size $L_y$ of the computational domain, for $\lambda=640\comp$, $\sigma=10$, and $\alpha=0.1$. The agreement between the black curve ($L_y=400\comp$) and the red line ($L_y=800\comp$) suggests that our results become insensitive to the transverse size of the box (i.e., they converge with respect to $L_y$), for boxes larger than $400\comp$. Generally speaking, we find that a good criterion for convergence is $L_y\gtrsim\lambda/2$.\footnote{We point out that this criterion is much more constraining than the requirement that the fastest growing tearing mode can fit in the width of the computational box. For relativistic reconnection, the most unstable wavelength is comparable to the current sheet thickness \citep{zenitani_07}, which in our setup is a few skin depths, much smaller than the stripe wavelength $\lambda=640\comp$.} When this condition is fulfilled, the growth of magnetic islands behind the fast shock is not artificially inhibited by the periodicity of our boundaries in the $y$ direction. Magnetic reconnection then  proceeds up to the point when islands from two neighboring current sheets will merge, which happens when their size is $\simeq\lambda/2$ (for $\alpha\lesssim0.1$).

On the other hand, for $L_{y}\lesssim 300\comp$ the growth and coalescence of reconnection islands within a given current sheet artificially stops when only one island is left in the sheet, and its size approaches $L_y$. At this point, a second shock forms, located behind the fast shock. For $L_y\gtrsim400\comp$, this would correspond to the hydrodynamic shock discussed in \S\ref{sec:shock}, where the striped structure of the flow is erased, and field energy is entirely transferred to the particles. Instead, in the case $L_{y}\lesssim 300\comp$, complete dissipation of the alternating fields is prohibited by the fact that the growth of islands is artificially inhibited by the transverse extent of the box. Here, the post-shock flow retains a striped structure, with regions of hot plasma separated by highly magnetized walls of cold particles. With decreasing $L_y$, less magnetic energy is transferred to the particles behind the fast shock, and the post-shock flow gets more dominated by magnetic field energy, with respect to particle kinetic energy (see the decrease in average particle energy with decreasing $L_y$, in the subpanel of \fig{specmy}). This also explains why in \fig{specmy} the normalization of the low-energy peak, populated by the cold particles in the high-field regions, grows with decreasing $L_y$, at the expense of the high-energy component of hot particles that gained energy from field dissipation. 

As the downstream region gets more magnetized with decreasing $L_y$, the main shock, which would be moving at $\beta_{\rm sh}\simeq1/3$ for $L_y\gtrsim400\comp$, propagates at a faster velocity, eventually catching up with the fast MHD shock, in the limit $L_y\ll\lambda$. In this regime, we recover the 1D results by \citet{petri_lyubarsky_07}. For the parameters employed here, the 1D model by  \citet{petri_lyubarsky_07} would predict negligible field dissipation, in agreement with our results for a very narrow box (yellow line for $L_y=12.5\comp$; see also the subpanel, for small values of $L_y$). However, as clarified by  \fig{specmy}, this is just an artificial consequence of the reduced dimensionality of their model, which cannot correctly capture the development of the tearing-mode instability, and the resulting growth and coalescence of magnetic islands. In a more realistic 2D scenario, complete field dissipation is achieved by the time the flow enters the hydrodynamic shock. This stresses that multi-dimensional simulations are essential for our understanding of shock-driven reconnection.

\vspace{-0.1in}
\section*{References}
\bibliography{sironi}

\end{document}